\documentclass[]{pasj01}

\usepackage{url}
\begin{document}
\SetRunningHead{Tomonari Michiyama et al.}{ASTE merger survey}
\title{Investigating the Relation between CO~(3--2) and Far Infrared Luminosities for Nearby Merging Galaxies Using ASTE}
                                                                                                                                                                                                                                                                                               
\author{%
   Tomonari \textsc{Michiyama},\altaffilmark{1,2}
   Daisuke \textsc{Iono},\altaffilmark{1,2}
   Kouichiro \textsc{Nakanishi}\altaffilmark{1,2}\\
   Junko \textsc{Ueda},\altaffilmark{3}
   Toshiki \textsc{Saito},\altaffilmark{4,2}
   Misaki \textsc{Ando},\altaffilmark{1,2}
   Hiroyuki \textsc{Kaneko},\altaffilmark{5,2}\\
   Takuji \textsc{Yamashita},\altaffilmark{6}
   Yuichi \textsc{Matsuda},\altaffilmark{1,2}
   Bunyo \textsc{Hatsukade},\altaffilmark{2}\\
   Kenichi \textsc{Kikuchi},\altaffilmark{2}
   Shinya \textsc{Komugi},\altaffilmark{7}
   and
   Takayuki \textsc{Muto}\altaffilmark{7}
   }
 \altaffiltext{1}{Department of Astronomical Science, SOKENDAI (The Graduate University of Advanced Studies), Mitaka, Tokyo 181-8588}
 \email{t.michiyama@nao.ac.jp}
 \altaffiltext{2}{National Astronomical Observatory of Japan, National Institutes of Natural Sciences, 2-21-1 Osawa, Mitaka, Tokyo, 181-8588}
 \altaffiltext{3}{Harvard-Smithsonian Center for Astrophysics, 60 Garden Street,Cambridge, MA 02138, USA}
 \altaffiltext{4}{Department of Astronomy, The University of Tokyo, 7-3-1 Hongo, Bunkyo-ku, Tokyo, 133-0033}
 \altaffiltext{5}{Nobeyama Radio Observatory, 462-2, Minamimaki, Minamisaku, Nagano, 384-1305}
 \altaffiltext{6}{Institute of Space and Astronautical Science,
Japan Aerospace Exploration Agency, 3-1-1 Yoshinodai, Chuo-ku, Sagamihara, Kanagawa 252-5210}
 \altaffiltext{7}{Division of Liberal Arts, Kogakuin University, 1-24-2, Nishi-Shinjuku, Shinjuku-ku, Tokyo 163-8677}

\KeyWords{Galaxies: evolution --- Galaxies: interactions --- Galaxies: starburst } 

\maketitle

\begin{abstract}
 We present the new single dish CO~(3--2) emission data obtained toward 19 early stage and 7 late stage nearby merging galaxies using the Atacama Submillimeter Telescope Experiment (ASTE).
Combining with the single dish and interferometric data of galaxies observed in previous studies, we investigate the relation between the CO~(3--2) luminosity ($L'_{\rm CO(3-2)}$) and the far Infrared luminosity ($L_{\rm FIR}$) in a sample of 29 early stage and 31 late stage merging galaxies, and 28 nearby isolated spiral galaxies.
We find that normal isolated spiral galaxies and merging galaxies have different slopes ($\alpha$) in the $\log L'_{\rm CO(3-2)}-\log L_{\rm FIR}$ plane ($\alpha\sim 0.79$ for spirals and $\sim 1.12$ for mergers).
The large slope ($\alpha>1$) for merging galaxies can be interpreted as an evidence for increasing Star Formation Efficiency (SFE=$L_{\rm FIR}/L'_{\rm CO(3-2)}$) as a function of $L_{\rm FIR}$.
Comparing our results with sub-kpc scale local star formation and global star-burst activity in the high-z Universe, we find deviations from the linear relationship in the $\log L'_{\rm CO(3-2)}-\log L_{\rm FIR}$ plane for the late stage mergers and high-z star forming galaxies.
Finally, we find that the average SFE gradually increases from isolated galaxies, merging galaxies, and to high-z submillimeter galaxies / quasi-stellar objects (SMGs/QSOs).  
By comparing our findings with the results from numerical simulations, we suggest; (1) inefficient star-bursts triggered by disk-wide dense clumps occur in the early stage of interaction and (2) efficient star-bursts triggered by central concentration of gas occur in the final stage. A systematic high spatial resolution survey of diffuse and dense gas tracers is a key to confirm this scenario.     
\end{abstract}

\section{Introduction}
\label{sec:intro}
Mergers between galaxies play a key role for a galaxy evolution since interacting galaxies have bridge or tailed structure which is not seen in isolated galaxies. In addition to the morphological evolution, an interaction between gas-rich progenitor galaxies condenses gas and triggers star-burst activities in the nuclear regions (e.g., \cite{Hopkins+05}) and throughout the extended disks \citep{SM+96}. In some instance, an Active Galactic Nuclei (AGN) is triggered by efficient feeding of the gas to the nuclear regions through a merger event (e.g., \cite{Hopkins+06,Nara+08,Hayward+14}).

Since molecular gas is the important ingredient for current and future star formations, observations of dense molecular gas are crucial for understanding the processes and timescales controlling star formation.
The far-infrared (FIR) luminosities and the luminosities of dense gas tracers such as CO~(3--2) and HCN~(1--0), whose critical densities are $\sim 10^{4-5}$ cm$^{-3}$ have observationally linear relation \citep{GS04a,GS04b,Nara+05} in local galaxies, and theoretical models basically suggest that linear relation is seen because dense gas is likely to be direct fuel for massive star formation \citep{Krumholz+07,Nara+08r}.  
In this regard, the CO~(3--2) emission line can be used to trace the moderately dense gas that is associated with star formation.
\citet{Iono+09} have shown that the integrated CO~(3--2) line luminosity ($L'_{\rm CO(3-2)}$) is correlated with the star formation rate (SFR) traced in FIR luminosities ($L_{\rm FIR}$), using Submillimeter array (SMA) observations of 14 local Ultra/luminous Infrared Galaxies (U/LIRGs) supplemented with the CO~(3--2) data obtained toward high-z sources  (see also \cite{Yao+03,Nara+05,Komugi+07,Bayet+09,Mao+10,Leech+10}).
This correlation was further investigated by \citet{Wilson+12} using more quiescent ``disk" galaxies in the local universe, finding that the Star Formation Efficiency (SFE=$L_{\rm FIR}/L'_{\rm CO(3-2)}$) of the quiescent galaxies is on average lower than those of the merging U/LIRGs. Moreover, \citet{Muraoka+16} suggest that the CO~(3--2) to FIR luminosity relation is universally applicable to different types of galaxies observed at different scales, from spatially resolved nearby galaxy disks to distant infrared (IR) luminous galaxies, within $\sim1$ dex scatter.
By comparing observations of low to high-{\itshape J} CO transitions, \citet{Greve+14} find that the slope in the $\log L'_{\rm CO}$--$\log L_{\rm FIR}$ plane is near unity for lower-{\itshape J} CO, but decreases gradually for higher-{\itshape J}  transitions (from 0.93 for {\itshape J} = 6--5 to 0.47 for {\itshape J} = 13--12). On the other hand, recent similar studies with larger sample sources \citep{Liu+15,Kamenetzky+15} suggest that the slopes are close to unity for high-{\itshape J} CO lines ({\itshape J} $>$ 4--3). 

From observations of CO emission in normal and star-burst galaxies at high-redshifts, it is found that isolated normal disk galaxies and star-burst systems follow a bimodal sequence, both in the molecular gas mass -- IR luminosity ($\log${\itshape M}$_{\rm H_2}$--$\log L_{\rm IR}$) plane, (or equivalently the molecular gas mass surface density -- SFR surface density plane; the Kenicutt-Schmidt relation, \cite{Komugi+05}) and the $\log L'_{\rm CO}$--$\log L_{\rm IR}$ plane, albeit the bimodality is less apparent in the latter \citep{Daddi+new,Daddi+old,Genzel+10}. Star formation in the disk galaxies is often regarded as the long-lasting mode, whereas galaxies in the star-burst sequence are experiencing a more rapid mode of star formation (higher SFE), possibly due to galaxies involved in a major merger.
In contrast, numerical simulations of merging galaxies provided by \citet{Powell+13} have shown that isolated and merging galaxies do not produce the bimodal relationship, and mergers are rather close to the sequence of disk galaxies \citep{Perret+14}.
Therefore, it is important to investigate the exact location and evolution of merging galaxies in the 
$\log L'_{\rm CO}$--$\log L_{\rm FIR}$ plane through a systematic observation toward early (before coalescence) and late stage (after coalescence) mergers with a large FIR range.
However, most of the observational studies had focused on late stage mergers that are bright in the FIR luminosity (U/LIRGs; e.g., \cite{Downes98}).
Therefore, we conducted a CO~(3--2) single dish survey using the Atacama Submillimeter Telescope Experiment (ASTE: \cite{Ezawa+04,Ezawa+08}) along the complete merger sequence with a wide FIR range ($10^{9}\LO<L_{\rm FIR}<10^{13}\LO$).

This paper is organized as follows.
We describe the details of ASTE observation and data analysis in section 2.
The relation between $L_{\rm FIR}$, $L'_{\rm CO(3-2)}$, and SFE in a variety of sources is investigated in section 3, and 
we conclude this paper in section 4. 
We adopted $H_{0}$ = 73 km s$^{-1}$ Mpc$^{-1}$, $\Omega_{\rm M}$ = 0.27, and $\Omega_{\Lambda}$ = 0.73 for all of the analysis throughout this paper.

\section{ASTE observation and data analysis}
\label{sec:sample}
\subsection{Sample selection}
\begin{figure}
 \begin{center}
  \includegraphics[width=8cm]{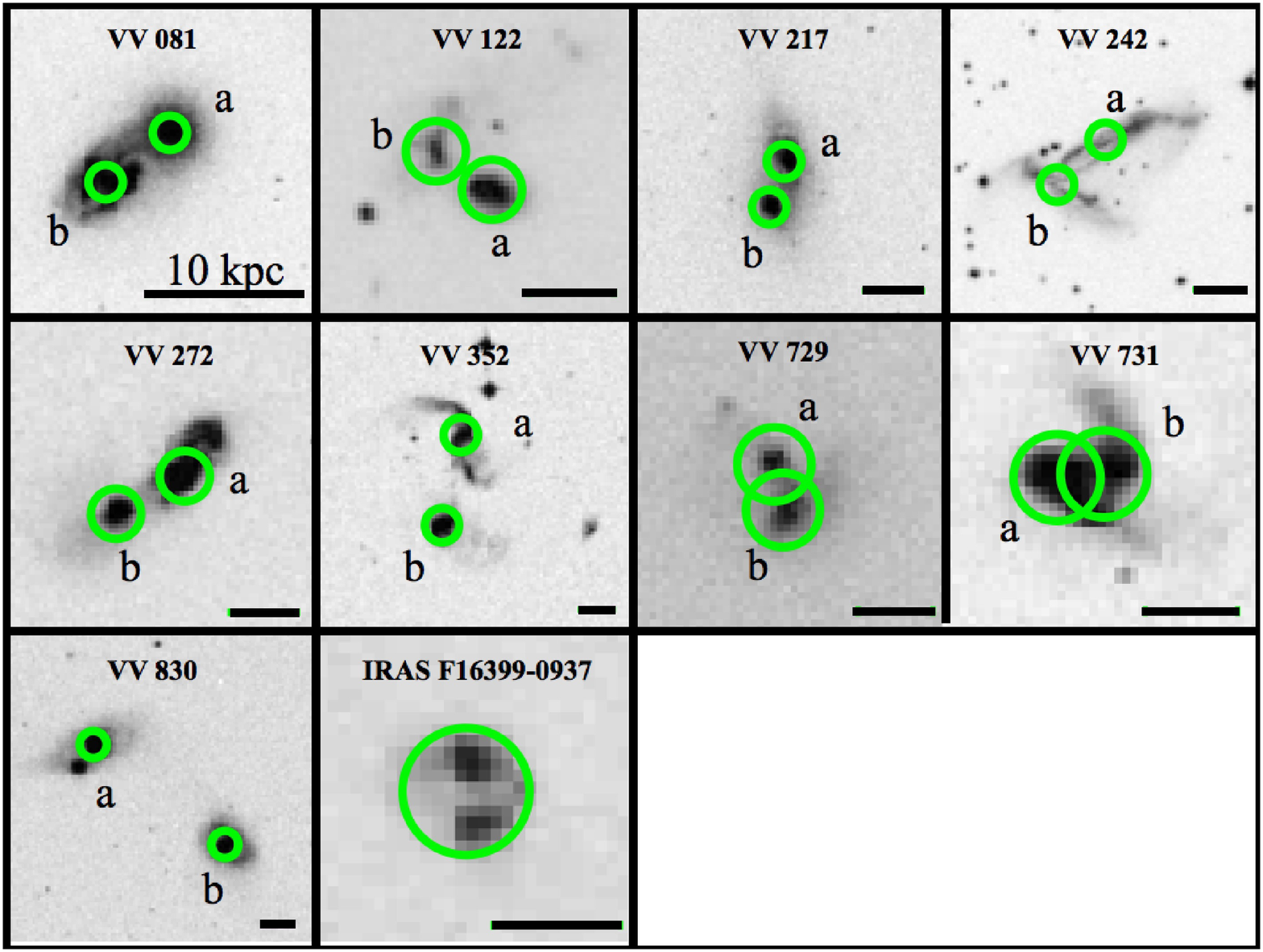} 
 \end{center}
\caption{The DSS blue band images of the early stage mergers. The green circles show the observing positions with ASTE 22$\arcsec$~beams. The lines on the bottom-right show the 10 kpc scale length.}\label{early}
\end{figure}

 We selected ``early stage merger" (double nuclei in optical image) from the VV-catalog \citep{VV+77,VV+01}, along the following criteria;
(a) both galaxies have measured {\itshape B}-band magnitudes, and the difference of the {\itshape B}-band magnitude between the two galaxies in a system is $<$ 3 mag (in order to select major mergers),
(b) the system is identified in the {\itshape IRAS} Revised Bright Galaxy Sample ($IRAS$ RBGS, \cite{Sanders+03}),
(c) the optical radial velocity is known for both sources,
and (d) the declination is $<$ 30 degrees. 
For the current observations, we selected nine pairs (VV~81, VV~217, VV~242, VV~272, VV~352, VV~729, VV~122, VV~830, and VV~731) out of the 40 galaxies which satisfy the criteria.
Additionally, we observed IRAS F16399-0937 as an early stage merging galaxy while the ASTE 22$\arcsec$~beam covers both nuclei separated by 3.4 kpc.   
The Digitized Sky Survey (DSS) images for these systems are shown in figure~$\ref{early}$, and the pointing positions and velocity information are presented in table~\ref{early-table}.

\begin{table*}
  \tbl{Early stage merging galaxy sample}{%
  \begin{tabular}{ccccc}
   \hline
   Source  & RA  &  DEC &  Vel.\footnotemark[$*$] \\
      & J2000 & J2000  & [km s$^{-1}$]\\
      \hline
VV 081a & \timeform{00h51m01.8s} & \timeform{-07D03'25"} & 1,750  \\
VV 081b & \timeform{00h51m04.4s} & \timeform{-07D03'56"} & 1,744  \\
VV 122a & \timeform{01h58m05.3s} & \timeform{+03D05'01"} & 5,431 \\
VV 122b & \timeform{01h58m06.6s} & \timeform{+03D05'15"} & 5,589 \\
VV 217a & \timeform{02h29m09.7s} & \timeform{-10D49'43"} & 4,686  \\
VV 217b & \timeform{02h29m10.3s} & \timeform{-10D50'10"} & 4,516  \\
VV 242a & \timeform{22h19m27.8s} & \timeform{+29D23'45"} & 4,569 \\
VV 242b & \timeform{22h19m30.0s} & \timeform{+29D23'17"} & 4,493\\
VV 272a & \timeform{00h06m27.0s} & \timeform{-13D24'58"} & 5,729  \\
VV 272b & \timeform{00h06m29.0s} & \timeform{-13D25'14"} & 5,717  \\
VV 352a & \timeform{00h18m50.1s} & \timeform{-10D21'42"} & 8,193  \\
VV 352b & \timeform{00h18m50.9s} & \timeform{-10D22'37"} & 8,125 \\
VV 729a & \timeform{03h41m10.5s} & \timeform{-01D18'10"} & 7,750  \\
VV 729b & \timeform{03h41m10.7s} & \timeform{-01D17'56"} & 7,592 \\
VV 731a & \timeform{23h18m22.6s} & \timeform{-04D24'58"} & 7,250 \\
VV 731b & \timeform{23h18m21.8s} & \timeform{-04D24'57"} & 7,380  \\
VV 830a & \timeform{00h42m52.8s} & \timeform{-23D32'28"} & 6,664 \\
VV 830b & \timeform{00h42m45.8s} & \timeform{-23D33'41"} & 6,787  \\
IRAS F16399-0937 & \timeform{16h42m40.2s} & \timeform{-09D43'14"} & 8,098\\
   \hline    \end{tabular}}\label{early-table}
\begin{tabnote}
      \par\noindent
      \footnotemark[$*$] Heliocentric Radial Velocity from NED.
\end{tabnote}
\end{table*}

\begin{figure}
 \begin{center}
  \includegraphics[width=8cm]{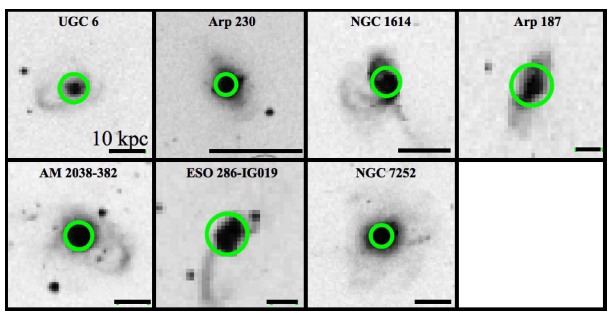} 
 \end{center}
\caption{The DSS blue band images of the late stage mergers. The green circles show the observing positions with ASTE 22\arcsec ~beams. The lines on the bottom-right show the 10 kpc scale length.}\label{late}
\end{figure}

Additionally, we selected three ``late stage merger" (single nucleus in optical image) from RBGS with declination $<$ 30 degrees (ESO~286-IG019, NGC~1614, and NGC~7252) and four (AM~2038-382, Arp~230, Arp~187, and UGC~6) with relatively low $L_{\rm FIR}$ ($< 10^{11}$ \LO) from \citet{Ueda+14}.
The DSS images for these systems are shown in figure~$\ref{late}$. The pointing positions and velocity information are presented in table~\ref{late-table}.

\begin{table*}
  \tbl{Late stage merging galaxy sample}{%
  \begin{tabular}{ccccc}
   \hline
   Source  & RA  &  DEC &  Vel.\footnotemark[$*$] \\
      & J2000 & J2000  & [km s$^{-1}$] \\
      \hline
UGC 6 & \timeform{00h03m09.6s} & \timeform{+21D57'39"} & 6,579\\
Arp 230 & \timeform{00h46m24.2s} & \timeform{-13D26'32"} & 1,720\\
NGC 1614 & \timeform{04h34m59.8s} & \timeform{-08D34'46"} & 4,778 \\
Arp 187 & \timeform{05h04m53.0s} & \timeform{-10D14'51"} & 12,095 \\
AM 2038-382 & \timeform{20h41m13.9s} & \timeform{-38D11'37"} & 6,092 \\
ESO 286-IG019 & \timeform{20h58m26.8s} & \timeform{-42D39'02"} & 12,890 \\
NGC 7252 & \timeform{22h20m44.7s} & \timeform{-24D40'42"} & 4,792\\
   \hline    \end{tabular}}\label{late-table}
\begin{tabnote}
      \par\noindent
      \footnotemark[$*$] Heliocentric Radial Velocity from NED.
\end{tabnote}
\end{table*}

\subsection{ASTE CO~(3--2) observation}
We observed the CO~(3--2) line emission toward the sample sources with ASTE during the observing seasons 2014B, 2015A, 2015B, and 2015C. The total observation time was $\sim$120 hours (including overhead) to observe all 26 individual galaxies. We conducted single point observations with the position switch mode. The main beam size is 22\arcsec ~at 345 GHz, and we specify the OFF position to be 5\arcmin ~away from the target sources. We used two types of receivers; CATS345 \citep{Inoue+08} in 2014B and DASH345 in 2015A, 2015B, and 2015C.
We used the 2048 MHz mode of WHSF \citep{Iguchi+08,Okuda+08} for the backend spectrometer (velocity resolution and coverage are 0.86 and 1750 km s$^{-1}$ at 350 GHz, respectively). R-sky calibration was carried out every 15 minutes, and the system noise temperature was typically 200--400 K. Absolute flux scales of the obtained spectra were calibrated by observing a standard source at least once a night. The average main beam efficiency is $\eta_{\rm mb}=0.57\pm0.08$. We checked the pointing of the telescope accuracy every 60--90 minutes and the errors were typically $<$ 3\arcsec.
The ASTE $22\arcsec$  beam corresponds to $\sim2.5$ kpc for the nearest target Arp 230 ($D_{\rm L}=19.3$ Mpc). Since the typical CO~(3--2) size of LIRGs is $0.3-3.1$ kpc \citep{Iono+09}, we assume that our ASTE $22\arcsec$ beam is large enough (at least comparable) to trace bulk of CO~(3--2) emission.

\subsection{Data Reduction and Analysis}
We use the ${\tt NEWSTAR}$ which is the software package developed at the Nobeyama Radio Observatory to process the raw data. Low quality spectra with a winding baseline are flagged by eye and only the high quality spectra are integrated. The flag rate ($\sim$ 30 -- 70 $\%$) strongly depends on the weather conditions. The spectra are then smoothed to a velocity resolution of 30 km s$^{-1}$ (for sources with CO~(3--2) detection) or 50 km s$^{-1}$ (for non-detection) with a boxcar function in order to improve the signal to noise ratio (S/N). Baselines are fitted with a polynomial function of degree one, but we used second and third orders in some cases with large baseline fluctuations. We successfully detected CO~(3--2) emission from 17 sources, and the final spectra are shown in figure~\ref{spectrum}. 

We derive the integrated CO~(3--2) properties for sources which show more than three continuous channels with positive $>$ 3$\sigma$ signal. The CO velocity-integrated intensity is derived using the following equation:
\begin{equation}
I_{\rm CO} = \int T_{\rm mb}dV = \int \frac{T_{\rm A}^{\star}}{\eta_{\rm mb}}dV
\end{equation}
where $I_{\rm CO}$ is in the unit of K km s$^{-1}$, $T_{\rm mb}$ is the main beam temperature in Kelvin, and $\eta_{\rm mb}$ is 0.47 - 0.71. The errors in $I_{\rm CO}$ for the value listed in table~\ref{detection} were calculated using
\begin{equation}
{I^{\rm err}_{\rm CO}} = \sigma_{\rm R.M.S.} \sqrt{(\Delta V_{\rm CO}\delta V)} 
\end{equation}
where $\sigma_{\rm R.M.S.}$ is the R.M.S. noise of $T_{\rm mb}$ in Kelvin, $\Delta V_{\rm CO}$ is the full line width in km s$^{-1}$ (the range between FWZI in table~\ref{detection}), and $\delta V$ is the velocity resolution in km s$^{-1}$ (30 km s$^{-1}$ for detected sources). 
The derived errors are in the range of 6--20 $\%$ of measured $I_{\rm CO}$. 
The $3\sigma$ upper limits of $I_{\rm CO}$ for non-detected sources were measured by assuming a gaussian profile,
\begin{equation}
I_{\rm CO}^{\rm upper} = \frac{\sqrt{2\pi}}{2}\sigma_V(\sigma_{\rm R.M.S.}\times 3)
\end{equation}
where $\sigma_V$ is the velocity dispersion of the emission line. We assumed $\sigma_V = 200$ km s$^{-1}$ (FWHM$\sim 500$ km s$^{-1}$) which is the maximam value of our CO~(3--2) detected sources (VV~272a).
The $L'_{\rm CO(3-2)}$ is calculated by using the following equation \citep{Solomon-Vanden+05},
\begin{equation}
L'_{\rm CO(3-2)}=23.5\Omega_{{\rm s}\star {\rm b}}D_{\rm L}^2I_{\rm CO}(1+z)^{-3}.
\end{equation}
The $L'_{\rm CO(3-2)}$ is given in K km s$^{-1}$ pc$^2$. The $\Omega_{{\rm s}\star {\rm b}}$ is the solid angle of the source convolved with the telescope beam in arcsec$^2$ assuming that CO is uniformly distributed in ASTE beam. The $D_{\rm L}$ is the luminosity distance in Mpc. We summarize the observational information in table~\ref{detection}.

\begin{figure*}
 \begin{center}
  \includegraphics[width=14cm]{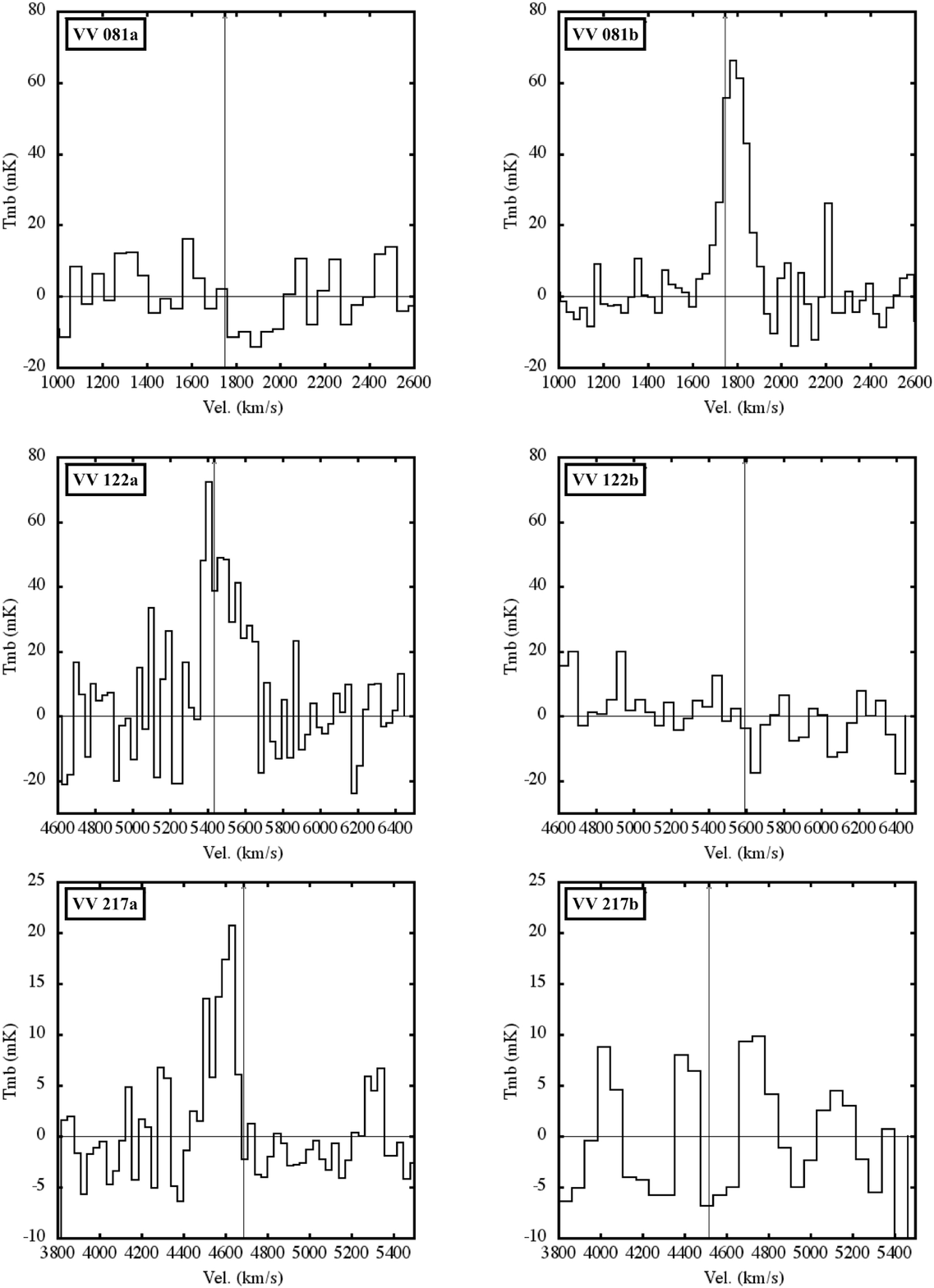}
  \caption{CO~(3--2) spectrum of our ASTE observation sample. The velocity resolution is 30 km s$^{-1}$ for detected sources and 50 km s$^{-1}$ for non-detected sources. The vertical line represents the systematic velocity from NED.}
 \end{center}
\end{figure*}

\setcounter{figure}{2}
\begin{figure*}
 \begin{center}
  \includegraphics[width=14cm]{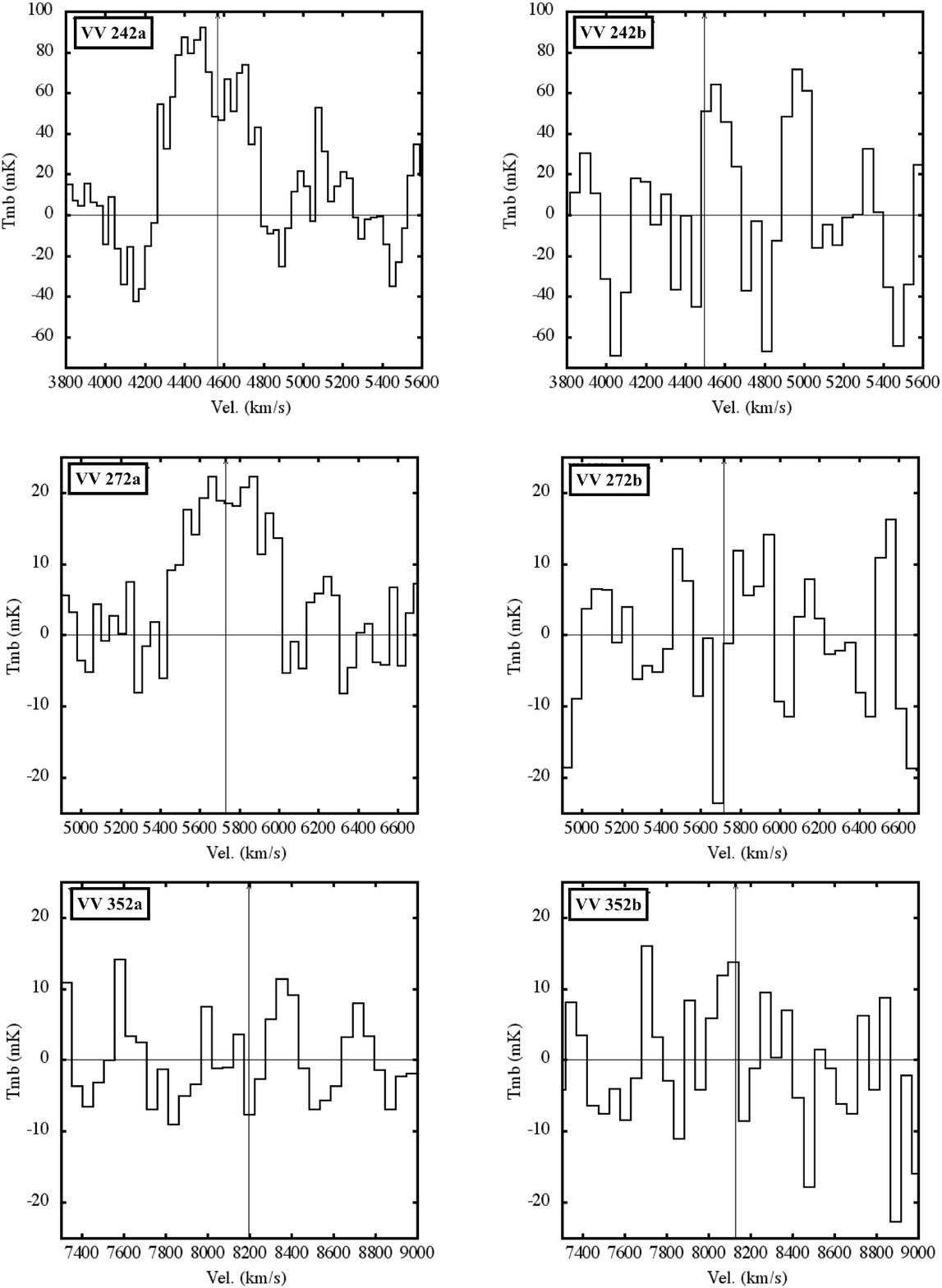}
  \caption{continued}
 \end{center}
\end{figure*}

\setcounter{figure}{2}
\begin{figure*}
 \begin{center}
  \includegraphics[width=14cm]{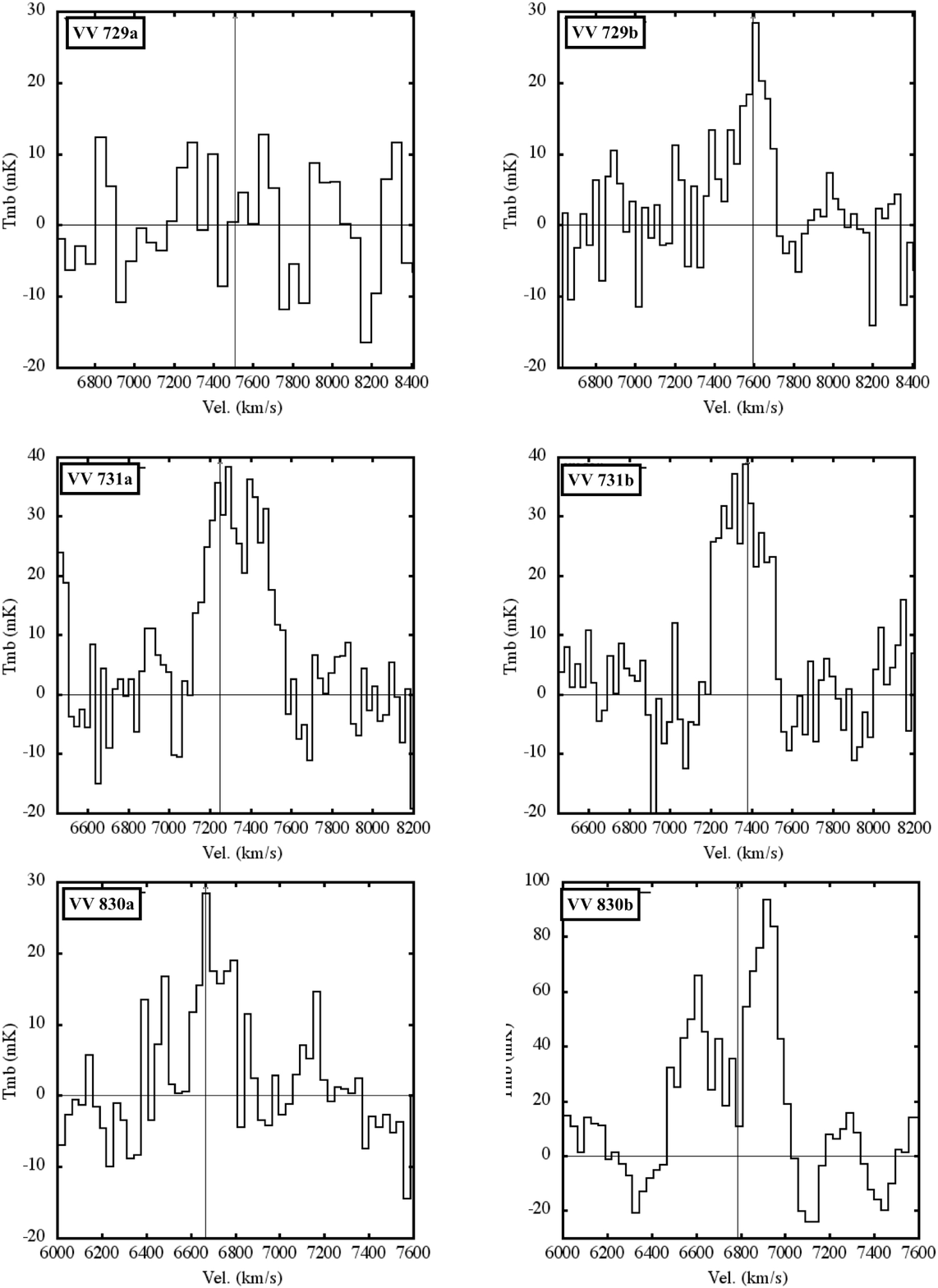}
  \caption{continued}
 \end{center}
\end{figure*}

\setcounter{figure}{2}
\begin{figure*}
 \begin{center}
  \includegraphics[width=14cm]{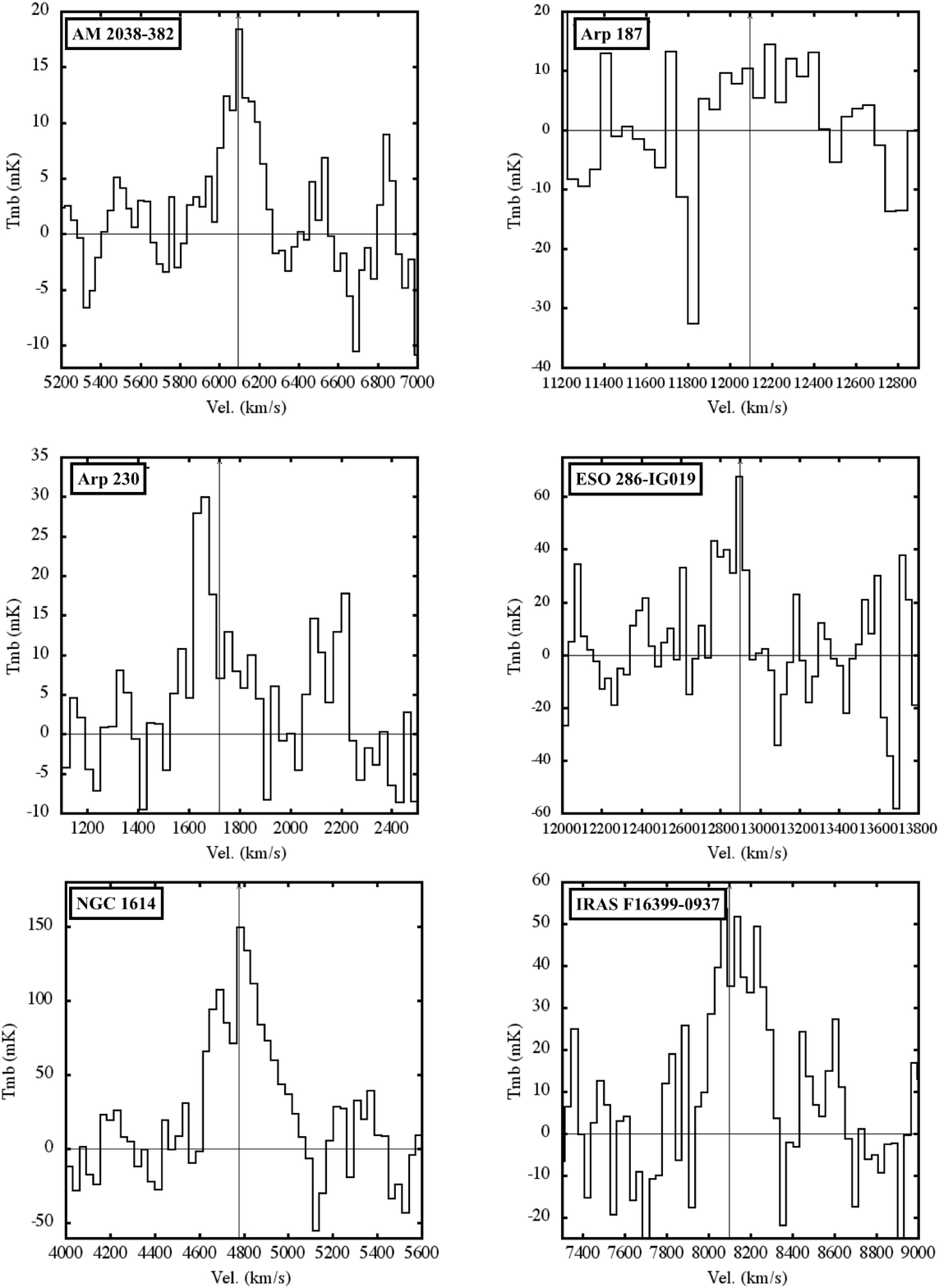}
  \caption{continued}
 \end{center}
\end{figure*}

\setcounter{figure}{2}
\begin{figure*}
 \begin{center}
  \includegraphics[width=14cm]{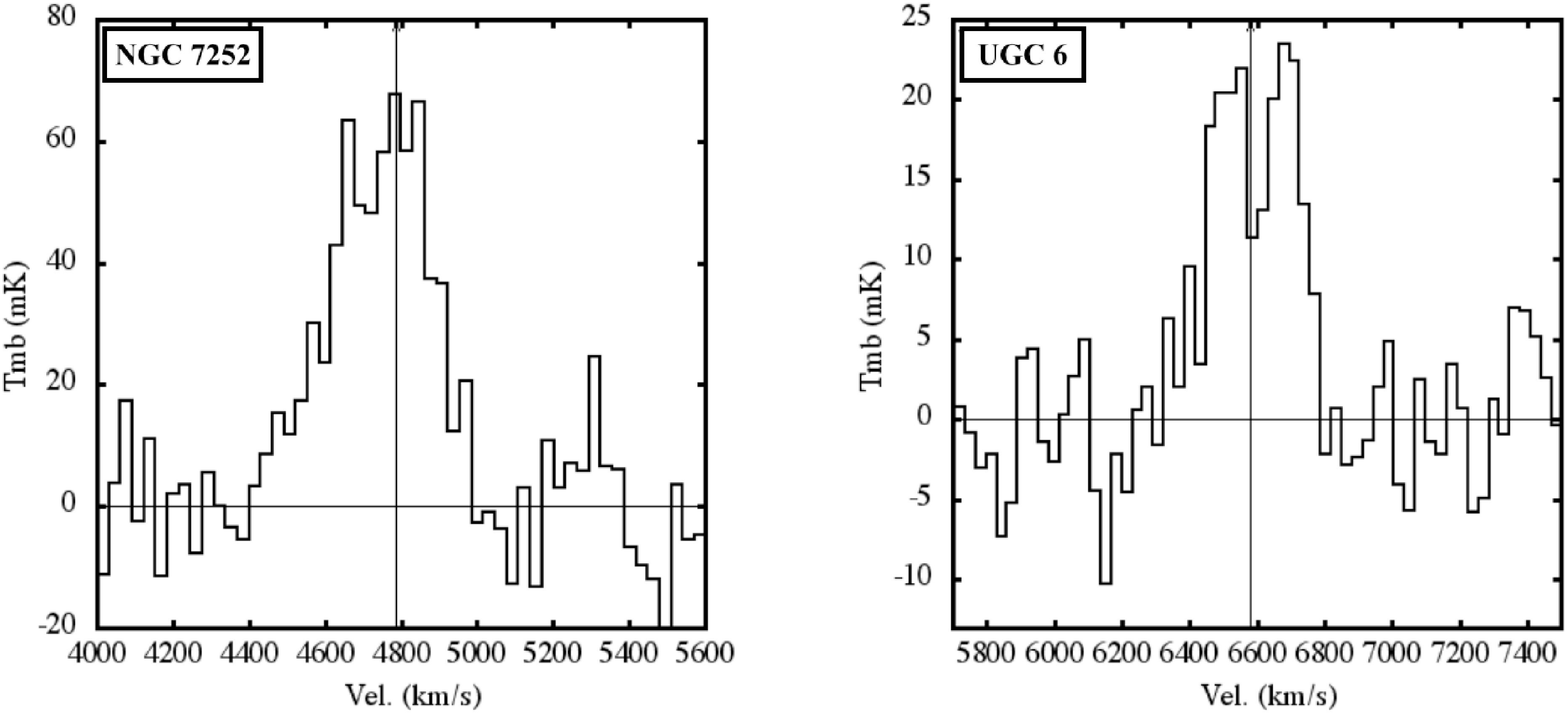} 
  \caption{continued}
\label{spectrum}
\end{center}
\end{figure*}

\begin{table*}
  \tbl{ASTE CO~(3--2) observation}{%
  \begin{tabular}{ccccccccccc}
      \hline
      source  & t$_{\rm integ}$\footnotemark[$*$] & $\sigma_{\rm R.M.S.}$\footnotemark[$\dagger$] & $T_{\rm peak}$\footnotemark[$\ddagger$] & S/N\footnotemark[$\S$] & FWZI\footnotemark[$\|$] & FWHM\footnotemark[$\#$] & $I_{\rm CO}$\footnotemark[$**$] & ${I^{\rm err}_{\rm CO}}$\footnotemark[$\dagger\dagger$]\\
      & & [mK] & [mK] & & [km s$^{-1}$] & [km s$^{-1}$] & [K km s$^{-1}$] & [K km s$^{-1}$]\\ 
      \hline
VV 081a & 14m20s & 8.4 & $<$~25.2 & - & - & - & $<$~6.32 & - \\
VV 081b & 28m15s & 7.1 & 66.3 & 9.3 & 1,620-1,920 & 130 & 9.13 & 0.67 \\
VV 122a & 6m50s & 13.0 & 72.3 & 5.6 & 5,360-5,660 & 150 & 11.85 & 1.23 \\
VV 122b & 31m30s & 6.9 & $<$~20.7 & - & - & - & $<$~5.19 & - \\
VV 217a & 129m00s & 3.3 & 20.7 & 6.3 & 4,430-4,670 & 100 & 2.45 & 0.28 \\
VV 217b & 105m25s & 5.6 & $<$~16.8 & - & - & - & $<$~4.21 & - \\
VV 242a & 8m50s & 20.5 & 92.3 & 4.5 & 4,270-4,780 & 410 & 32.83 & 2.54 \\
VV 242b & 24m40s & 37.9 & $<$~113.7 & - & - & - & $<$~28.50 & - \\
VV 272a & 49m05s & 4.7 & 22.3 & 4.7 & 5,440-5,890 & 500 & 9.41 & 0.55 \\
VV 272b & 25m10s & 8.7 & $<$~26.1 & - & - & - & $<$~6.54 & - \\
VV 352a & 29m00s & 6.0 & $<$~18.0 & - & - & - & $<$~4.51 & - \\
VV 352b & 118m00s & 7.9 & $<$~23.7 & - & - & - & $<$~5.94 & - \\
VV 729a & 11m30s & 7.6 & $<$~22.8 & - & - & - & $<$~5.72 & - \\
VV 729b & 37m50s & 5.7 & 28.4 & 5.0 & 7,350-7,710 & 150 & 4.83 & 0.59 \\
VV 731a & 39m40s & 6.3 & 38.3 & 6.1 & 7,120-7,630 & 330 & 11.31 & 0.78 \\
VV 731b & 40m50s & 8.5 & 38.9 & 4.6 & 7,150-7,600 & 300 & 8.96 & 0.99 \\
VV 830a & 17m30s & 6.3 & 28.3 & 4.5 & 6,590-6,800 & 170 & 3.90 & 0.50 \\
VV 830b & 22m30s & 11.7 & 93.4 & 8.0 & 6,470-7,010 & 160 & 25.60 & 1.49 \\
IRAS F16399-0937 & 25m00s & 13.7 & 53.6 & 3.9 & 7,940-8,330 & 300 & 12.15 & 1.48 \\
AM 2038-382 & 90m30s & 3.8 & 18.4 & 4.8 & 5,840-6,260 & 180 & 2.82 & 0.43 \\
Arp 187 & 33m20s & 10.2 & $<$~30.6 & - & - & - & $<$~7.67 & - \\
Arp 230 & 95m40s & 6.7 & 30.0 & 4.5 & 1,530-1,890 & 90 & 4.25 & 0.70 \\
ESO 286-IG019 & 10m10s & 14.5 & 67.4 & 4.6 & 12,760-12,940 & 50 & 7.89 & 1.07 \\
NGC 1614 & 4m30s & 22.4 & 149.5 & 6.7 & 4,620-5,120 & 150 & 35.30 & 2.60 \\
NGC 7252 & 19m50s & 9.2 & 67.7 & 7.4 & 4,400-4,970 & 300 & 20.51 & 1.20 \\
UGC 6 & 55m10s & 4.1 & 23.5 & 5.7 & 6,330-6,780 & 160 & 6.61 & 0.48 \\
      \hline
    \end{tabular}}\label{detection}
\begin{tabnote}
      \par\noindent
      \footnotemark[$*$] Total on-source time after flagging bad baseline spectra.
      \par\noindent
      \footnotemark[$\dagger$] Root mean square noise level measured with ${\tt NEWSTAR}$ in main beam temperature.
      \par\noindent
      \footnotemark[$\ddagger$] Peak temperature of CO~(3--2) emission in main beam temperature.
      \par\noindent
      \footnotemark[$\S$] Signal to noise ratio. $T_{\rm peak}/\sigma_{\rm R.M.S.}$.
      \par\noindent
      \footnotemark[$\|$] The full width zero intensity.
      \par\noindent
      \footnotemark[$\#$] Full width half maximum for emission spectrum.
      \par\noindent
      \footnotemark[$**$] CO~(3--2) intensity in the unit of K km s$^{-1}$.
      \par\noindent
      \footnotemark[$\dagger\dagger$] The error of CO~(3--2) intensity in the unit of K km s$^{-1}$.
\end{tabnote}
\end{table*}

\subsection{Supplementary data}
We supplement the sample by adding 10 early stage merging galaxies obtained using SMA \citep{Wilson+08} and James Clark Maxwell Telescope (JCMT) \citep{Leech+10}, yielding a combined sample of 29 early stage mergers with a wide range of FIR luminosity ($10^9~L_{\odot}< L_{\rm FIR} < 10^{12}~L_{\odot}$). In addition, we use 24 late stage merging galaxies compiled from the literature \citep{Wilson+08,Leech+10}, yielding a combined sample of 31 late stage merging galaxies with a wide range of FIR luminosity ($10^9~L_{\odot}< L_{\rm FIR} < 10^{13}~L_{\odot}$). 
We have visually checked the DSS blue band image and classified the merger stage. 
We classified the system as an early stage merger when the two progenitor galaxies are separated by $>10\arcsec$, and the rest to be late stage mergers.

We use the JCMT CO~(3--2) maps obtained toward 28 nearby isolated spiral galaxies \citep{Wilson+12} as a control sample. This sample allows us to compare the star formation relation among early stage mergers, late stage mergers, and non-merging systems. The isolated spiral galaxies are on average closer ($D_L < 40$ Mpc) than the merging galaxies ($D_L > 60$ Mpc).

\begin{figure}
 \begin{center}
  \includegraphics[width=8cm]{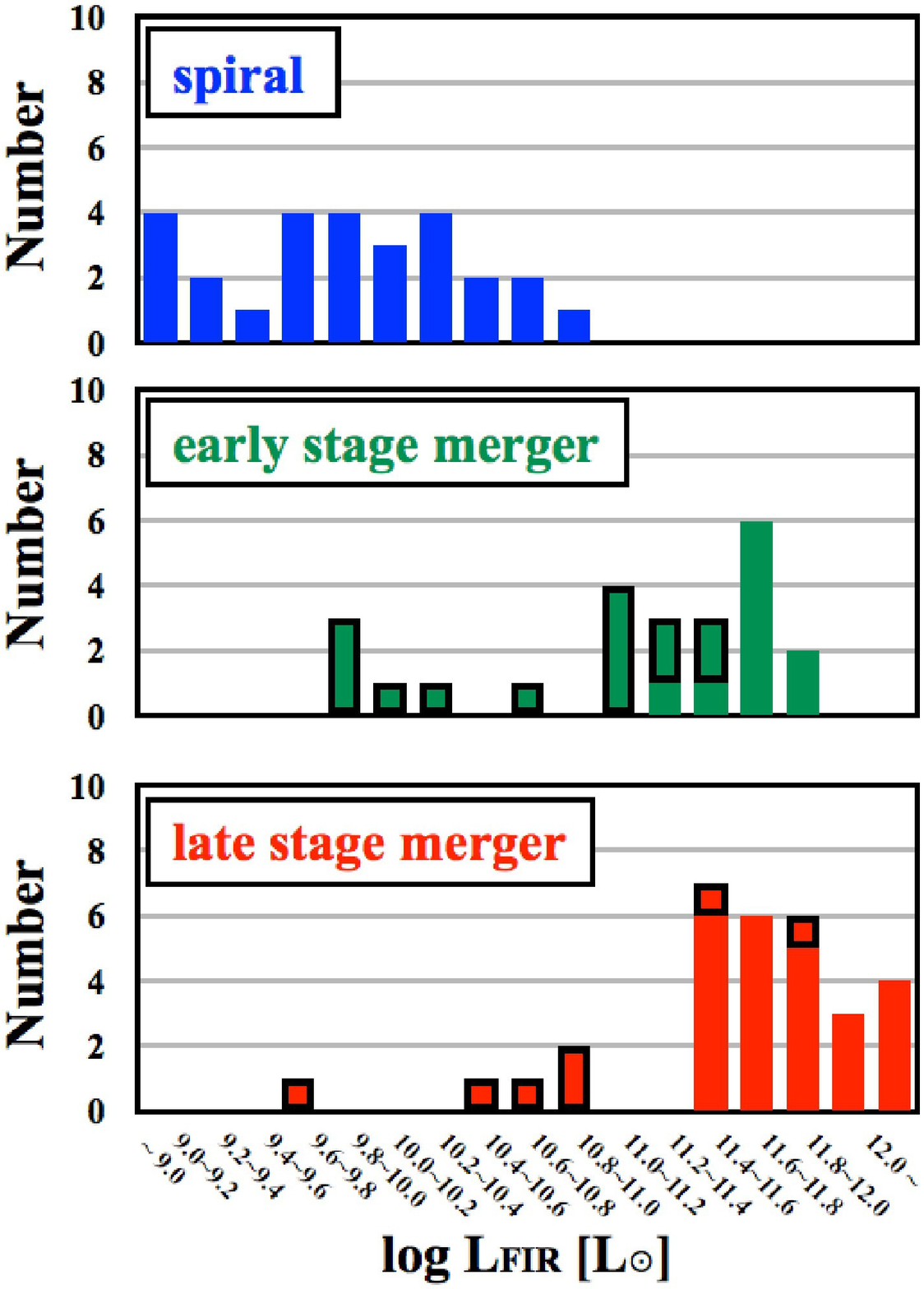} 
 \end{center}
\caption{Histogram of the FIR luminosities of the three samples. The green, red, and blue bars show early-stage mergers, late stage mergers, and isolated spirals, respectively. The black boxes show sources which we observed with ASTE, and the others are galaxies from the literature.}\label{hist}
\end{figure}

The histogram of the FIR luminosity of the sample sources is shown in figure~$\ref{hist}$. The number of early stage merging galaxies in the low FIR luminosity range ($10^9\LO<L_{\rm FIR}<10^{11}\LO$) has especially increased in this work.
Finally, we note that there are other single-dish CO~(3--2) observations in the literature obtained toward nearby galaxies (e.g.,\cite{Yao+03,Vila+03,Komugi+07,Nara+08,Mao+10,Papa+12,Greve+14}). Since most of the data were taken with a smaller beam with respect to the size of the galaxy, or the derivation of the FIR luminosity is different from ours, we will not use these data in our comparative analysis.

\subsection{FIR Luminosity derived by {\itshape AKARI}}
We use the 90 $\mu$m and 140 $\mu$m photometries in the {\itshape AKARI} FIS Bright Source Catalog ({\itshape AKARI}/FIS BSC) \citep{Yamamura+10} to derive the FIR luminosity ($L_{\rm FIR}$). 
In order to estimate $L_{\rm FIR}$ in a consistent manner toward all sample sources, we use the DARTS catalog match system\footnote{\url{http://www.darts.isas.jaxa.jp/astro/akari/}} 
to obtain the FIR counterpart in the {\itshape AKARI}/FIS BSC. 
We use the formulation described in \citet{Takeuchi+10}. 
The error $L_{\rm FIR}^{\rm err}$ is measured from {\itshape AKARI} $90\,\mathrm{\mu m}$ and $140\,\mathrm{\mu m}$ flux error ($\sim$ 5 $\%$ and 10 $\%$ respectively).
The spatial resolution of {\itshape AKARI} is too coarse to estimate FIR flux separately for the two galaxies in the early stage mergers. One way to separate the contribution is to scale the FIR luminosity using the ratio of the radio continuum emission, assuming that the radio to FIR correlation holds true in these galaxies \citep{Yun+01}. Another way is to scale through the {\itshape K}-band flux ratio, again assuming a constant ratio between the SFR (derived from the FIR luminosity) and the stellar mass (derived from {\itshape K}-band luminosity). By using the 1.5 GHz images obtained from the Karl G. Jansky Very Large Array (VLA) archive, we find that six of the early stage mergers have radio emission only from one galaxy in a system (VV~081, VV~122, VV~242, VV~272, and VV~352).  For these sources, we assume that all of the FIR emission arises from a galaxy with the radio detection. If the radio emission is detected from both galaxies, we distribute the total FIR luminosity according to the flux ratio of the 1.5 GHz emission ($a:b=38:62$ for VV 731 and $39:61$ for VV 830). Finally, we use the flux ratio of 2MASS \citep{Cutri+03,Jarrett+03} $2.2\,\mathrm{\mu m}$ ($a:b=55:45$ for VV 217 and $63:37$ for VV 729) for galaxies with no VLA detection from either of the galaxy in the pair. 

\subsection{Derivation of the Star Formation Efficiency (SFE)}
We assumed that the CO~(3--2) emission line measures the total fuel for molecular gas which is the direct material of star formation in a galaxy, and $L_{\rm FIR}$ arises mostly from dusty star-forming regions. 
Therefore, the comparison between $L'_{\rm CO(3-2)}$ and $L_{\rm FIR}$ connects the amount of available molecular gas mass to the amount of current massive star formation.
We discuss the effect of AGNs in section \ref{secAGN}.
We define SFE by taking the ratio between the FIR luminosity and the CO~(3--2) luminosity,
\begin{equation}
{\rm SFE}~[\LO~{\rm (K~km~s^{-1}~pc^2)^{-1}}] =L_{\rm FIR}/L'_{\rm CO(3-2)}.
\end{equation}
 While the ratio between the SFR and the molecular gas mass (i.e., SFR/$M_{\rm H_2}$) is often used as the SFE, here we use $L_{\rm FIR}/L'_{\rm CO(3-2)}$ to avoid introducing additional ambiguities through the variations and uncertainties pertaining to the CO to H$_2$ conversion factor and the CO excitation. We list $L_{\rm FIR}$, $L'_{\rm CO(3-2)}$, and SFE for each source in table~\ref{Allsample}.

\begin{longtable}{ccccc}
  \caption{Nearby Sample Sources}\label{Allsample}
  \hline              
 Source & $L_{\rm FIR}$\footnotemark[$*$] & $L'_{\rm CO(3-2)}$ & SFE & ref.\footnotemark[$\dagger$]  \\
  & [$10^{10}$~\LO] & [$10^8~$K km s$^{-1}$] & [\LO (K km s$^{-1}$ pc$^2$)$^{-1}$] &  \\
   \hline 
\endfirsthead
  \hline
 Source & $L_{\rm FIR}$\footnotemark[$*$] & $L'_{\rm CO(3-2)}$ & SFE & ref.\footnotemark[$\dagger$]  \\
 & [$10^{10}$~\LO] & [$10^8$~K km s$^{-1}$] & [\LO (K km s$^{-1}$ pc$^2$)$^{-1}$] &  \\
   \hline
\endhead
  \hline
  \multicolumn{5}{c}{
\begin{minipage}{10cm}\vspace*{3mm} 
{\footnotesize 
            \par\noindent
      \footnotemark[$*$] FIR luminosity based on the {\itshape AKARI}/FIS BSC (Section $2.3$). The ``--" sings means VLA non-detected galaxies of early stage merging galaxies. 
      \par\noindent
      \footnotemark[$\dagger$] The reference of $L'_{\rm CO(3-2)}$, \citet{Wilson+08}, \citet{Leech+10} or \citet{Wilson+12} 
      \par\noindent
      \footnotemark[$\ddagger$] Those galaxies are not listed in the {\itshape AKARI}/FIS BSC. We use FIR luminosities in referred paper (ref.) instead.
} \end{minipage} 
} 
\endfoot
  \hline
    \multicolumn{5}{c}{
\begin{minipage}{10cm}\vspace*{3mm} 
{\footnotesize 
            \par\noindent
      \footnotemark[$*$] FIR luminosity based on 90~$\mu$m and 140 $\mu$m of the {\itshape AKARI}/FIS BSC (Section $2.3$). The ``--" sings means VLA non-detected galaxies of early stage merging galaxies. 
      \par\noindent
      \footnotemark[$\dagger$] The reference of $L'_{\rm CO(3-2)}$, \citet{Wilson+08}, \citet{Leech+10} or \citet{Wilson+12} 
      \par\noindent
      \footnotemark[$\ddagger$] Those galaxies are not listed in the {\itshape AKARI}/FIS BSC. We use FIR luminosities in referred paper (ref.) instead.
      } \end{minipage} 
}
\endlastfoot
  \hline
\multicolumn{5}{c}{----- early stage merging galaxies -----}\\
VV 081a & - & $<$~0.20 & - & This work \\
VV 081b & 0.62$\pm$0.03 & 0.39$\pm$0.03 & 159$\pm$14 & This work \\
VV 122a & 6.33$\pm$0.30 & 6.55$\pm$0.68 & 97$\pm$11 & This work \\
VV 122b & - & $<$~3.23 & - & This work \\
VV 217a & 0.61$\pm$0.06 & 1.02$\pm$0.12 & 60$\pm$9 & This work \\
VV 217b & 0.46$\pm$0.05 & $<$~1.71 & $>$~38 & This work \\
VV 242a & 7.06$\pm$0.41 & 12.29$\pm$0.95 & 57$\pm$6 & This work \\
VV 242b & - & $<$~10.92 & - & This work \\
VV 272a & 3.27$\pm$0.33$\ddagger$ & 5.67$\pm$0.33 & 58$\pm$7 & This work \\
VV 272b & - & $<$~4.17 & - & This work \\
VV 352a & - & $<$~6.08 & - & This work \\
VV 352b & 17.94$\pm$0.34 & $<$~7.87 & $>$~326 & This work \\
VV 729a & 1.58$\pm$0.16$\ddagger$ & $<$~7.18 & $>$~31 & This work \\
VV 729b & 0.93$\pm$0.09$\ddagger$ & 5.51$\pm$0.68 & 29$\pm$5 & This work \\
VV 731a & 6.78$\pm$0.34 & 11.40$\pm$0.79 & 59$\pm$5 & This work \\
VV 731b & 11.36$\pm$0.58 & 9.08$\pm$1.00 & 125$\pm$15 & This work \\
VV 830a & 7.20$\pm$0.45 & 3.26$\pm$0.42 & 221$\pm$31 & This work \\
VV 830b & 11.72$\pm$0.73 & 22.30$\pm$1.30 & 53$\pm$4 & This work \\
IRAS F16399-0937 & 24.25$\pm$0.90 & 16.58$\pm$2.02 & 146$\pm$19 & This work \\
Arp 299 & 29.99$\pm$2.00 & 12.59$\pm$0.28 & 238$\pm$17 & \cite{Wilson+08} \\
NGC 5257/8 & 13.27$\pm$1.27 & 31.62$\pm$1.93 & 42$\pm$5 & \cite{Wilson+08} \\
NGC 5331 & 35.79$\pm$1.42 & 25.12$\pm$0.37 & 142$\pm$6 & \cite{Wilson+08} \\
Arp 236 & 27.24$\pm$1.06 & 50.12$\pm$0.99 & 54$\pm$2 & \cite{Leech+10} \\
UGC 2369 & 31.73$\pm$1.08 & 15.85$\pm$1.78 & 200$\pm$23 & \cite{Leech+10} \\
IRAS 03359+1523 & 27.18$\pm$1.73 & 12.59$\pm$2.56 & 216$\pm$46 & \cite{Leech+10} \\
Arp 55 & 41.09$\pm$1.82 & 63.10$\pm$5.32 & 65$\pm$6 & \cite{Leech+10} \\
Arp 238 & 39.81$\pm$4.04 & 12.59$\pm$2.09 & 316$\pm$61 & \cite{Leech+10} \\
Arp 302 & 43.37$\pm$1.63 & 50.12$\pm$2.04 & 87$\pm$5 & \cite{Leech+10} \\
NGC 6670 & 26.28$\pm$1.54 & 31.62$\pm$2.41 & 83$\pm$8 & \cite{Leech+10} \\
\multicolumn{5}{c}{----- late stage merging galaxies -----}\\
AM 2038-382 & 2.31$\pm$0.19 & 2.03$\pm$0.31 & 114$\pm$20 & This work \\
Arp 187 & 2.51$\pm$0.25$\ddagger$ & $<$~24.05 & $>$~15 & This work \\
Arp 230 & 0.30$\pm$0.02 & 0.18$\pm$0.03 & 168$\pm$31 & This work \\
ESO 286-IG019 & 57.07$\pm$3.05 & 25.63$\pm$3.46 & 223$\pm$32 & This work \\
NGC 1614 & 24.25$\pm$1.19 & 16.43$\pm$1.21 & 148$\pm$13 & This work \\
NGC 7252 & 4.35$\pm$0.33 & 8.61$\pm$0.50 & 51$\pm$5 & This work \\
UGC 6 & 5.90$\pm$0.28 & 5.29$\pm$0.38 & 111$\pm$10 & This work \\
NGC 2623 & 26.52$\pm$1.12 & 10.00$\pm$0.08 & 265$\pm$11 & \cite{Wilson+08} \\
Mrk 231 & 146.46$\pm$4.28 & 25.12$\pm$0.65 & 583$\pm$23 & \cite{Wilson+08} \\
Arp 193 & 37.13$\pm$1.57 & 25.12$\pm$0.42 & 148$\pm$7 & \cite{Wilson+08} \\
Mrk 273 & 85.47$\pm$2.58 & 31.62$\pm$1.00 & 270$\pm$12 & \cite{Wilson+08} \\
NGC 6240 & 43.92$\pm$1.75 & 79.43$\pm$1.70 & 55$\pm$3 & \cite{Wilson+08} \\
IRAS 17208-0014 & 158.54$\pm$10.52 & 50.12$\pm$3.46 & 316$\pm$30 & \cite{Wilson+08} \\
IRAS 00057+4021 & 24.12$\pm$1.35 & 15.85$\pm$0.83 & 152$\pm$12 & \cite{Leech+10} \\
IRAS 01077-1707 & 34.45$\pm$1.53 & 25.12$\pm$2.45 & 137$\pm$15 & \cite{Leech+10} \\
III Zw 35 & 25.94$\pm$0.75 & 12.59$\pm$1.50 & 206$\pm$25 & \cite{Leech+10} \\
Mrk 1027 & 18.17$\pm$0.71 & 50.12$\pm$2.04 & 36$\pm$2 & \cite{Leech+10} \\
IRAS 02483+4302 & 47.03$\pm$2.45 & 12.59$\pm$1.54 & 374$\pm$50 & \cite{Leech+10} \\
IRAS 04232+1436 & 117.76$\pm$5.77 & 100.00$\pm$12.00 & 118$\pm$15 & \cite{Leech+10} \\
IRAS 10039-3338 & 30.74$\pm$1.37 & 15.85$\pm$0.62 & 194$\pm$12 & \cite{Leech+10} \\
IRAS 10190+1322 & 79.00$\pm$19.20 & 50.12$\pm$5.35 & 158$\pm$42 & \cite{Leech+10} \\
IRAS 10565+2448 & 83.43$\pm$3.49 & 39.81$\pm$1.26 & 210$\pm$11 & \cite{Leech+10} \\
IRAS 13001-2339 & 24.01$\pm$1.93 & 15.85$\pm$0.54 & 151$\pm$13 & \cite{Leech+10} \\
NGC 5256 & 20.87$\pm$1.27 & 25.12$\pm$2.66 & 83$\pm$10 & \cite{Leech+10} \\
Mrk 673 & 21.54$\pm$1.64 & 19.95$\pm$2.96 & 108$\pm$18 & \cite{Leech+10} \\
IRAS  14348-1447 & 167.73$\pm$13.15 & 100.00$\pm$10.31 & 168$\pm$22 & \cite{Leech+10} \\
Mrk 848 & 46.69$\pm$1.85 & 25.12$\pm$1.39 & 186$\pm$13 & \cite{Leech+10} \\
NGC 6090 & 20.77$\pm$0.85 & 39.81$\pm$1.55 & 52$\pm$3 & \cite{Leech+10} \\
IRAS 17132+5313 & 53.61$\pm$1.97 & 25.12$\pm$5.02 & 213$\pm$43 & \cite{Leech+10} \\
IRAS 20010-2352 & 37.06$\pm$2.13 & 31.62$\pm$2.70 & 117$\pm$12 & \cite{Leech+10} \\
II Zw 96 & 45.26$\pm$1.77 & 31.62$\pm$1.98 & 143$\pm$11 & \cite{Leech+10} \\
\multicolumn{5}{c}{----- spiral galaxies -----}\\
NGC 0628 & 0.35$\pm$0.07 & 0.52$\pm$0.10 & 68$\pm$19 & \cite{Wilson+12} \\
NGC 0925 & 0.20$\pm$0.04 & 0.09$\pm$0.02 & 227$\pm$68 & \cite{Wilson+12} \\
NGC 2403 & 0.12$\pm$0.02 & 0.17$\pm$0.03 & 72$\pm$19 & \cite{Wilson+12} \\
NGC 2976 & 0.05$\pm$0.01 & 0.05$\pm$0.01 & 105$\pm$27 & \cite{Wilson+12} \\
NGC 3031 & 0.25$\pm$0.05 & 0.10$\pm$0.04 & 251$\pm$109 & \cite{Wilson+12} \\
NGC 3034 & 3.98$\pm$0.80 & 3.98$\pm$0.04 & 100$\pm$20 & \cite{Wilson+12} \\
NGC 3049 & 0.25$\pm$0.05 & 0.13$\pm$0.03 & 200$\pm$58 & \cite{Wilson+12} \\
NGC 3184 & 0.40$\pm$0.08 & 1.00$\pm$0.15 & 40$\pm$10 & \cite{Wilson+12} \\
NGC 3198 & 0.40$\pm$0.08 & 0.63$\pm$0.10 & 63$\pm$16 & \cite{Wilson+12} \\
NGC 3351 & 0.50$\pm$0.10 & 0.50$\pm$0.06 & 100$\pm$23 & \cite{Wilson+12} \\
NGC 3521 & 1.00$\pm$0.20 & 2.00$\pm$0.13 & 50$\pm$11 & \cite{Wilson+12} \\
NGC 3627 & 1.58$\pm$0.32 & 3.16$\pm$0.17 & 50$\pm$10 & \cite{Wilson+12} \\
NGC 3773 & 0.03$\pm$0.01 & 0.02$\pm$0.00 & 126$\pm$40 & \cite{Wilson+12} \\
NGC 3938 & 0.79$\pm$0.16 & 1.26$\pm$0.20 & 63$\pm$16 & \cite{Wilson+12} \\
NGC 4236 & 0.02$\pm$0.00 & 0.05$\pm$0.01 & 32$\pm$11 & \cite{Wilson+12} \\
NGC 4254 & 3.16$\pm$0.63 & 7.94$\pm$0.52 & 40$\pm$8 & \cite{Wilson+12} \\
NGC 4321 & 2.51$\pm$0.50 & 5.01$\pm$0.49 & 50$\pm$11 & \cite{Wilson+12} \\
NGC 4450 & 0.13$\pm$0.03 & 0.10$\pm$0.03 & 126$\pm$43 & \cite{Wilson+12} \\
NGC 4559 & 0.25$\pm$0.05 & 0.20$\pm$0.04 & 126$\pm$37 & \cite{Wilson+12} \\
NGC 4569 & 1.00$\pm$0.20 & 2.00$\pm$0.17 & 50$\pm$11 & \cite{Wilson+12} \\
NGC 4579 & 0.63$\pm$0.13 & 0.79$\pm$0.14 & 79$\pm$21 & \cite{Wilson+12} \\
NGC 4625 & 0.03$\pm$0.01 & 0.01$\pm$0.00 & 200$\pm$64 & \cite{Wilson+12} \\
NGC 4631 & 1.26$\pm$0.25 & 1.58$\pm$0.08 & 79$\pm$16 & \cite{Wilson+12} \\
NGC 4736 & 0.50$\pm$0.10 & 0.50$\pm$0.04 & 100$\pm$21 & \cite{Wilson+12} \\
NGC 4826 & 0.63$\pm$0.13 & 1.00$\pm$0.05 & 63$\pm$13 & \cite{Wilson+12} \\
NGC 5033 & 1.58$\pm$0.32 & 2.51$\pm$0.33 & 63$\pm$15 & \cite{Wilson+12} \\
NGC 5055 & 1.00$\pm$0.20 & 2.00$\pm$0.17 & 50$\pm$11 & \cite{Wilson+12} \\
NGC 5194 & 1.58$\pm$0.32 & 5.01$\pm$0.19 & 32$\pm$6 & \cite{Wilson+12} \\
\end{longtable}

\section{Results and discussions}
\subsection{Relation between $L'_{\rm CO(3-2)}$ and $L_{\rm FIR}$}\label{sec:FIR-CO}
\begin{figure*}
 \begin{center}
  \includegraphics[width=14cm]{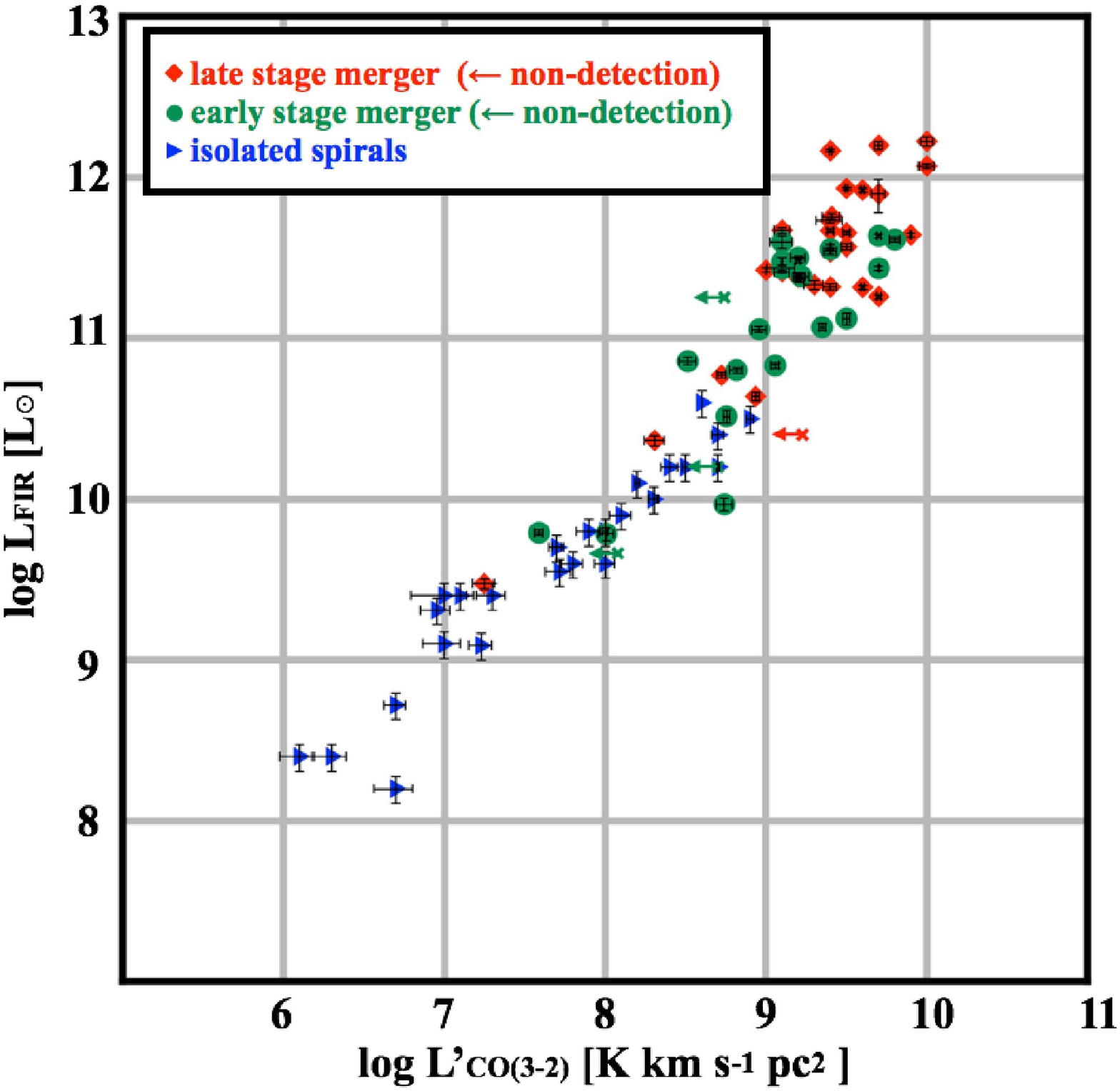} 
 \end{center}
\caption{Relation between $L'_{\rm CO(3-2)}$ and $L_{\rm FIR}$. The symbols colored in green, red, and blue represent early stage mergers, late stage mergers, and isolated spiral galaxies, respectively. The arrows represent the 3$\sigma$ upper limit of the CO~(3--2) luminosity for the ASTE non-detected sources. Galaxies that are not detected in both the CO~(3--2) and 1.5 GHz continuum emission are not shown here.}
\label{FIR-CO}
\end{figure*}

A linear relation (slope $\alpha\sim1$, $\log L_{\rm FIR}= \alpha \log L'_{\rm CO(3-2)}+\beta$) between $\log L'_{\rm CO(3-2)}$ and $\log L_{\rm FIR}$ is suggested in previous studies (e.g., \cite{Greve+14}, and references therein). Here we investigate the variation in the $L'_{\rm CO(3-2)}$--$L_{\rm FIR}$ correlation as a function of merger stage, and the results are presented in figure~\ref{FIR-CO}.  We use the nonlinear least-squares Marquardt-Levenberg algorithm with the \verb|fit| command of gnuplot and derive the following relation. 
\begin{equation}\label{all-equ}  
\begin{array}{l}
({\rm for}\ {\rm all}\ {\rm sample}\ {\rm sources})\\
\log L_{\rm FIR}=1.10{\scriptstyle \pm0.07}\ \log L'_{\rm CO(3-2)} +1.15{\scriptstyle \pm0.40}\\
{\rm r=0.95}
\end{array}
\end{equation}
The $L_{\rm FIR}^{\rm err}$ and ${I^{\rm err}_{\rm CO}}$ are reflected on the relative weight of each data point before determining the weighted sum of squared residuals. We use the correlation coefficient $r_{xy}=\frac{\Sigma(x_i-\overline{x})(y_i-\overline{y})}{\sqrt{\Sigma(x_i-\overline{x})^2(y_i-\overline{y})^2}}$ (where $x$ is $\log L'_{\rm CO(3-2)}$ and $y$ is $\log L_{\rm FIR}$) to evaluate the strength of the correlation between two variables. 
The derived slope of $\alpha = 1.10\pm0.07$ is consistent with the one derived by \citet{Kamenetzky+15} ($\alpha = 1.18\pm0.03$), but it is slightly larger than the linear relation obtained for U/LIRGs ($\alpha = 0.99\pm0.04$, \cite{Greve+14}). 

The spiral galaxies with low luminosity ($L_{\rm FIR}\sim10^{9}\LO$) appear to have systematically lower $L'_{\rm CO(3-2)}$ offsets.
\citet{Wilson+12} suggest the systematic offset may be due to lower metallicities in less luminous (i.e., lower mass) galaxy. In this case the CO luminosity may systematically underestimate the molecular hydrogen gas mass.
While the relative deficiency of the CO~(3--2) luminosity in the lower luminosity regime generally causes the slope to be steeper, these galaxies carry relatively low weight due to large uncertainties and hence the contribution to the overall fit is relatively small. 

\subsection{The bimodal star formation relation}
\citet{Daddi+new} suggest the existence of a bimodality in the CO -- FIR correlation (disk sequence and star-burst sequence), from their compilation of data from  local spiral galaxies, U/LIRGs, high-z SMGs and normal high-z galaxies (BzK galaxies).
Our result presented in figure~\ref{FIR-CO} does not show the evidence of the bimodality. 
This contradiction might be due to biased sampling and the various CO transitions used in \citet{Daddi+new}.
The galaxies on the star-burst sequence are biased toward extremely efficient star-forming galaxies. \citet{Saintonge+12} also suggest the lack of the bimodality using a less biased sample.
The CO transitions that \citet{Daddi+new} used range from CO~(1--0) to CO~(9--8), which makes the interpretation extremely complex due to the uncertainties in the excitation condition.

\subsection{Evolution of mergers in the $\log L'_{\rm CO(3-2)}$--$\log L_{\rm FIR}$ plane}
\label{two-sequence}

\begin{figure}
 \begin{center}
  \includegraphics[width=8cm]{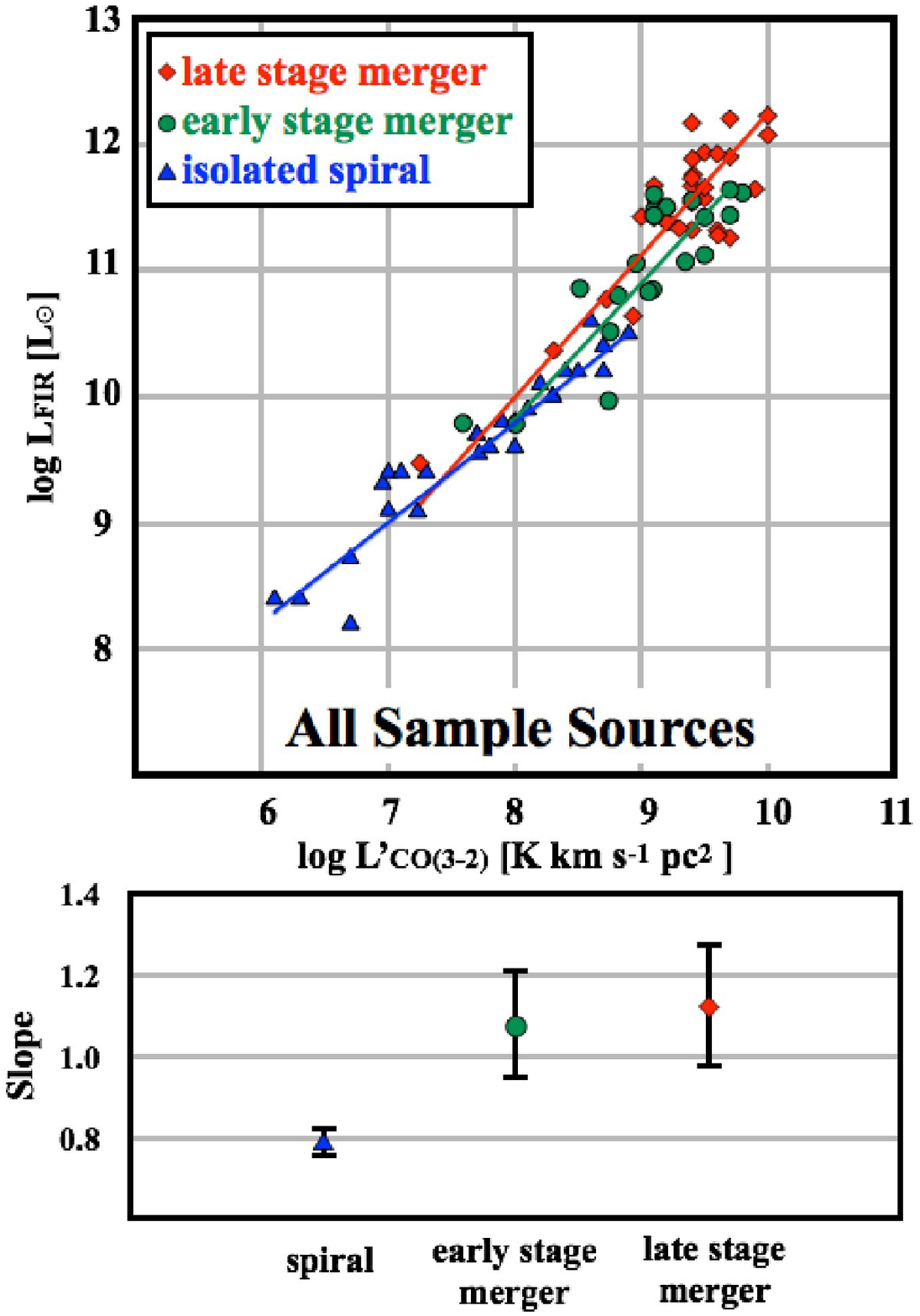} 
 \end{center}
\caption{(\textit{Top}) Relation between $L'_{\rm CO(3-2)}$ and $L_{\rm FIR}$. The best-fit functions are shown in colored lines. (\textit{Bottom}) the difference of slope between early stage mergers, late stage mergers, and isolated spiral galaxies with the fitting errors are presented. Galaxies that are not detected in the CO~(3--2) are not shown here.}
\label{slope}
\end{figure}

We find that isolated spiral galaxies, early and late stage mergers have different slope in the $\log L'_{\rm CO(3-2)}$--$\log L_{\rm FIR}$ plane (figure~\ref{slope}).
We performed a least square fitting separately towards late stage mergers, early stage mergers, and spiral galaxies (figure~\ref{slope}). The results are
\begin{equation}  
\begin{array}{l}
({\rm for}\ {\rm spirals})\\
\log L_{\rm FIR}=0.79{\scriptstyle \pm0.04}\ \log L'_{\rm CO(3-2)} + 3.50{\scriptstyle \pm0.24}\\
{\rm r=0.95}
\end{array}
\end{equation}
\begin{equation}  
\begin{array}{l}
({\rm for}\ {\rm early}\ {\rm stage}\ {\rm mergers})\\
\log L_{\rm FIR}=1.08{\scriptstyle \pm0.14}\ \log L'_{\rm CO(3-2)} + 1.18{\scriptstyle \pm0.61}\\
{\rm r=0.87}
\end{array}
\end{equation}
\begin{equation}  
\begin{array}{l}
({\rm for}\ {\rm late}\ {\rm stage}\ {\rm mergers})\\
\log L_{\rm FIR}=1.12{\scriptstyle \pm0.16}\ \log L'_{\rm CO(3-2)} + 1.00{\scriptstyle \pm0.65}\\
{\rm r=0.88}
\end{array}
\end{equation}
\begin{equation}  
\begin{array}{l}
({\rm for}\ {\rm all}\ {\rm mergers})\\
\log L_{\rm FIR}=1.10{\scriptstyle \pm0.07}\ \log L'_{\rm CO(3-2)} + 1.15{\scriptstyle \pm0.40}\\
{\rm r=0.85}
\end{array}
\end{equation}
We argue that the slopes are significantly different between isolated galaxies and mergers (as a whole), and appear to increase as a function of merger stage ($\alpha= 0.79\pm0.04,\ 1.08\pm0.14,\ 1.12\pm0.16$ for spirals, early stage mergers and late stage mergers, respectively). However, the evolution of the slope between the early and the late stage merger is not conclusive because of the large error bars.

\begin{figure}
 \begin{center}
  \includegraphics[width=8cm]{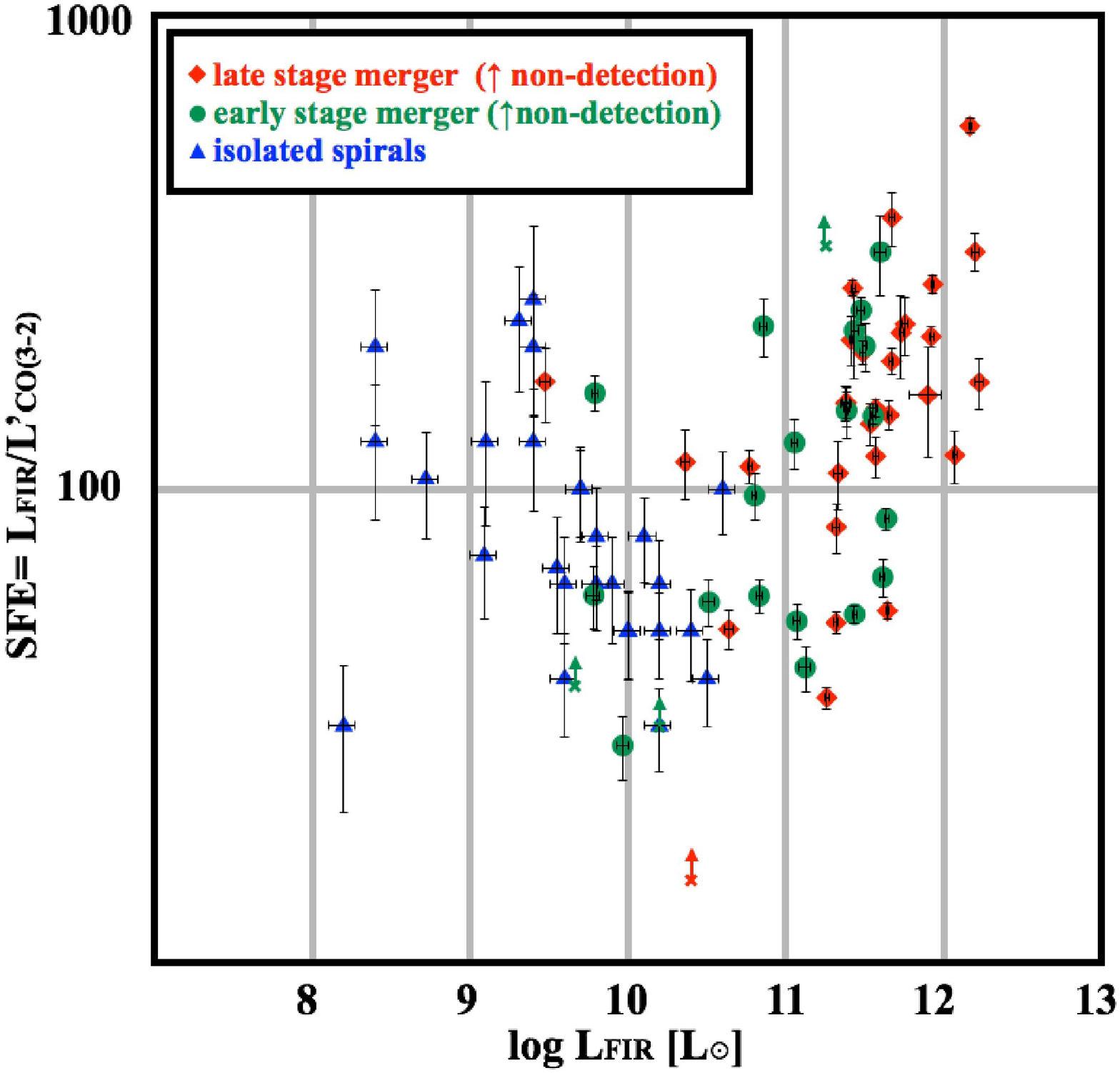} 
 \end{center}
\caption{Relation between $L_{\rm FIR}$ and SFE.The symbols and the colors are the same as figure \ref{FIR-CO}. The arrows represent the lower limit of SFE for the ASTE non-detected sources. Galaxies that are not detected in both the CO~(3--2) and 1.5 GHz continuum emission are not shown here.}
\label{FIR-SFE}
\end{figure}

The important finding from the above analysis is that merging galaxies have a slope larger than unity ($\alpha>1.0$) whereas spiral galaxies have $\alpha<1.0$. 
This is reflected in the relation between $L_{\rm FIR}$ and SFE (figure~\ref{FIR-SFE}), where the SFE of isolated spiral galaxies have a decreasing trend as a function of $L_{\rm FIR}$ whereas the mergers show the contrary. A possible positive correlation (r=0.55) for merging galaxies in figure~\ref{FIR-SFE} and the apparent increase of SFE from early to late stage merger are evident, which suggests that the efficiency of converting gas to stars is higher in late stage mergers.
We note that \citet{Iono+09} did not find a strong correlation between $L_{\rm FIR}$ and SFE, and we attribute this to the lack of galaxies with $10^{10} \LO$ $<$ $L_{\rm FIR}$ $<$ $10^{11} \LO$. 

\subsection{The effect of AGNs to the $L'_{\rm CO(3-2)}$--$L_{\rm FIR}$ relation}
\label{secAGN}
\begin{figure}
 \begin{center}
  \includegraphics[width=8cm]{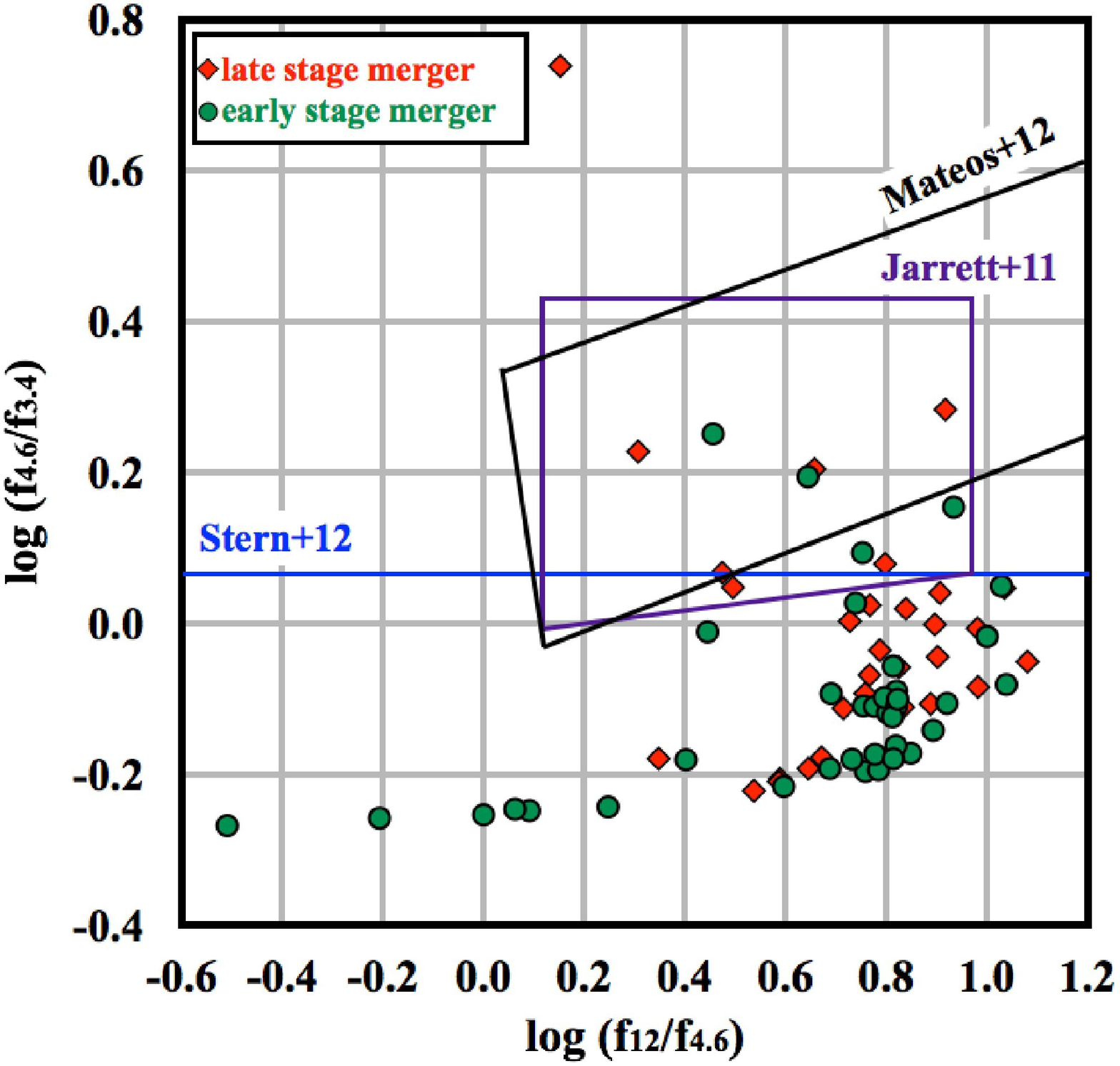} 
 \end{center}
\caption{WISE color-color diagram. Green and red symbols represent the early and late stage mergers. The region enclosed by the lines show the criteria for AGN candidates; \cite{Jarrett+11} (purple), \cite{Stern+12} (blue), and \cite{Mateos+12} (black).}
\label{wise}
\end{figure}

The contribution of the AGNs to the FIR luminosity can not be ignored for some sources \citep{SM+96}, and the exact fraction varies from source to source \citep{Armus+07}.
For example, \citet{Davies+04} show that the nuclear star-burst contribution in Mrk~231 is 25-40$\%$ of the bolometric luminosity. In contrast, \citet{Ichikawa+14} suggest that the energy contribution from AGNs against the total IR luminosity is typically only ~20$\%$ even in ULIRGs, suggesting that the majority of the IR luminosity originates from star-burst activities. We use the {\itshape Wide-field Infrared Survey Explorer} ({\itshape WISE}, \cite{Wright+10,Cutri+11}) 22\arcsec aperture photometry color of our sample sources (table~\ref{WISEcolor})
to select the AGN candidates, and  we identify nine sources based on the commonly used classification scheme \citep{Jarrett+11,Mateos+12,Stern+12} (figure~ \ref{wise}).

\begin{figure}
 \begin{center}
  \includegraphics[width=8cm]{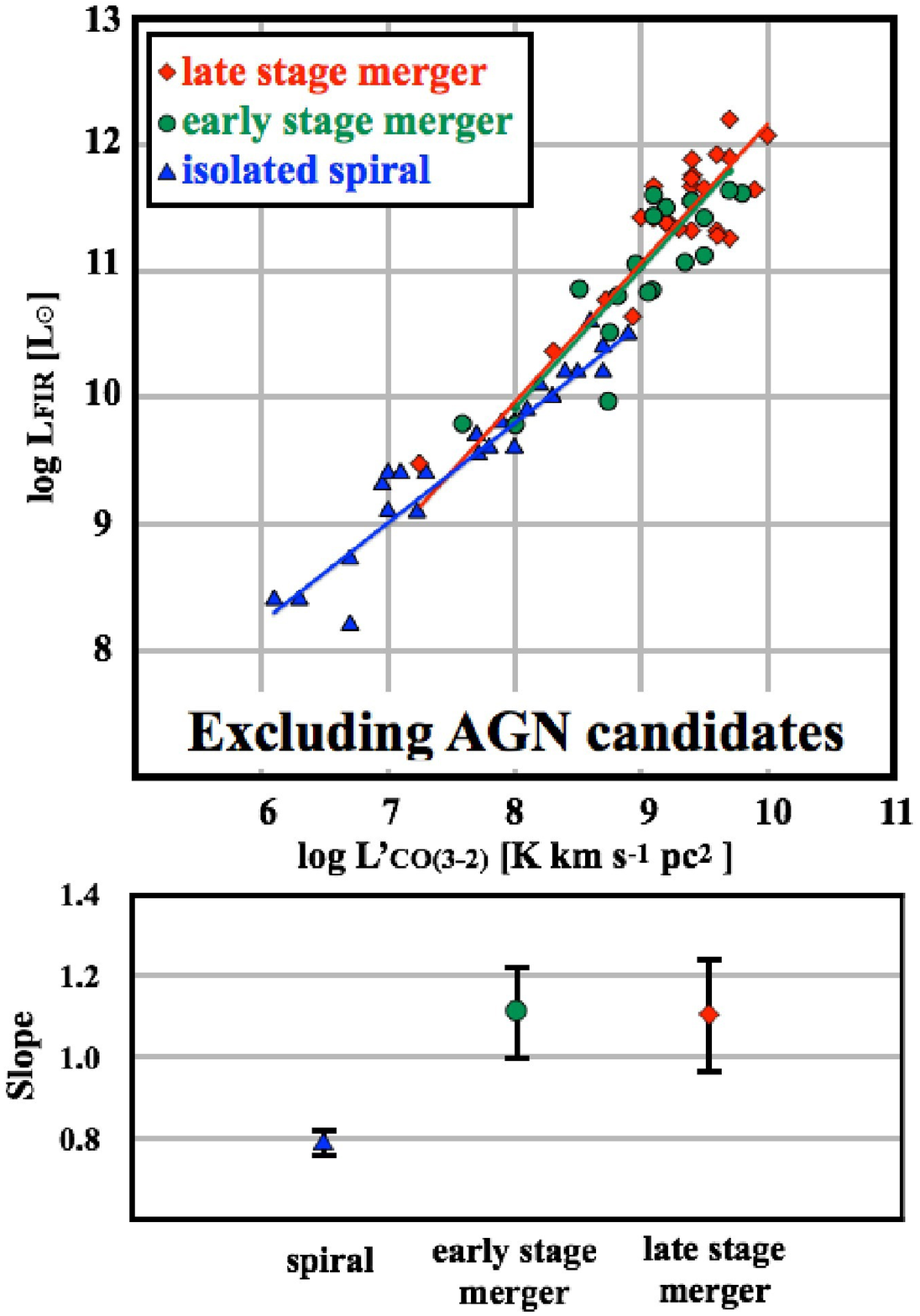} 
 \end{center}
\caption{(\textit{Top}) Relation between $L'_{\rm CO(3-2)}$ and $L_{\rm FIR}$ excluding AGN candidates. The best-fit functions are shown in colored lines. (\textit{Bottom}) the difference of slope between early stage mergers, late stage mergers, and isolated spiral galaxies with the fitting errors are presented. Galaxies that are not detected in the CO~(3--2) are not shown here.}
\label{slope2}
\end{figure}

Excluding these sources, the relation between $L'_{\rm CO(3-2)}$ and $L_{\rm FIR}$ becomes, 
\begin{equation}  
\begin{array}{l}
\log L_{\rm FIR}=1.11{\scriptstyle \pm0.12}\ \log L'_{\rm CO(3-2)} + 1.05{\scriptstyle \pm0.55}\\
{\rm r=0.87}
\end{array}
\end{equation}
for the early stage mergers, and
\begin{equation}
\begin{array}{l}
\log L_{\rm FIR}=1.10{\scriptstyle \pm0.15}\ \log L'_{\rm CO(3-2)} + 1.15{\scriptstyle \pm0.63}\\
{\rm r=0.88}
\end{array}
\end{equation}
for the late stage mergers (figure~\ref{slope2}).
There is little difference between the relation here and those derived in section \ref{two-sequence} (figure~\ref{slope}), and we conclude that the effect of the AGN to the $\log L'_{\rm CO(3-2)}$ and $\log L_{\rm FIR}$ relation is negligible. 

\begin{longtable}{ccc}
  \caption{WISE color}\label{WISEcolor}
  \hline              
 Source & log(f$_{12}$/f$_{4.6}$)& log(f$_{4.6}$/f$_{3.4}$)\\
   \hline 
\endfirsthead
  \hline
 Source & log(f$_{12}$/f$_{4.6}$)& log(f$_{4.6}$/f$_{3.4}$)\\
   \hline
\endhead
  \hline
  \multicolumn{3}{c}{
\begin{minipage}{10cm}\vspace*{3mm} 
{\footnotesize
      \par\noindent
      \footnotemark[$*$] AGN candidates.
} \end{minipage} 
} 
\endfoot
  \hline
    \multicolumn{3}{c}{
\begin{minipage}{10cm}\vspace*{3mm} 
{\footnotesize 
      \par\noindent
      \footnotemark[$*$] AGN candidates.
} \end{minipage} 
}
\endlastfoot
  \hline
\multicolumn{3}{c}{----- early stage merging galaxies -----}\\
VV 081a & 0.31 & 0.54 \\
VV 081b & 5.73 & 0.64 \\
VV 122a & 7.81 & 0.72 \\
VV 122b & 7.05 & 0.67 \\
VV 217a & 1.23 & 0.56 \\
VV 217b & 0.62 & 0.55 \\
VV 242a & 5.68 & 0.78 \\
VV 242b & 3.94 & 0.61 \\
VV 272a & 5.39 & 0.66 \\
VV 272b & 1.15 & 0.57 \\
VV 352a & 6.08 & 0.64 \\
VV 352b & 10.92 & 0.83 \\
VV 729a & 1.00 & 0.56 \\
VV 729b & 2.52 & 0.66 \\
VV 731a & 6.51 & 0.88 \\
VV 731b & 5.48 & 1.06 \\
VV 830a & 2.78 & 0.97 \\
VV 830b & 5.97 & 0.78 \\
Arp 299W\footnotemark[$*$] & 0.93 & 0.15 \\
Arp 299E\footnotemark[$*$] & 0.46 & 0.25 \\
NGC 5257W & 0.78 & -0.17 \\
NGC 5257E & 0.82 & -0.16 \\
NGC 5331N & 0.69 & -0.19 \\
NGC 5331S & 0.80 & -0.12 \\
Arp 236W\footnotemark[$*$] & 0.64 & 0.19 \\
Arp 236E\footnotemark[$*$] & 0.75 & 0.09 \\
UGC 2369N & 0.25 & -0.24 \\
UGC 2369S & 1.00 & -0.02 \\
Arp 55E & 0.80 & -0.10 \\
Arp 55W & 0.82 & -0.11 \\
Arp 238E & 1.03 & 0.05 \\
Arp 238W & 0.81 & -0.12 \\
Arp 302N & 0.69 & -0.09 \\
Arp 302S & 0.81 & -0.18 \\
NGC 6670E & 0.92 & -0.11 \\
NGC 6670W & 0.82 & -0.10 \\
\multicolumn{3}{c}{----- late stage merging galaxies -----}\\
AM 2038-382 & 0.54 & -0.22 \\
Arp 187 & 0.59 & -0.21 \\
Arp 230 & 0.59 & -0.21 \\
ESO 286-IG019 & 0.92 & 0.28 \\
IRAS F16399-0937 & 0.82 & -0.09 \\
NGC 1614 & 1.08 & -0.05 \\
NGC 7252 & 0.65 & -0.19 \\
UGC 6 & 0.50 & 0.05 \\
NGC 2623 & 0.77 & 0.02 \\
Mrk 231\footnotemark[$*$] & 0.31 & 0.23 \\
Arp 193 & 0.90 & -0.00 \\
Mrk 273\footnotemark[$*$] & 0.66 & 0.20 \\
NGC 6240 & 0.73 & 0.00 \\
IRAS 17208-0014 & 0.84 & 0.02 \\
IRAS 00057+4021 & 0.78 & -0.19 \\
IRAS 01077-1707 & 0.89 & -0.11 \\
III ZW35 & 0.79 & -0.04 \\
Mrk 1027 & 0.83 & -0.11 \\
IRAS 02483+4302 & 0.35 & -0.18 \\
IRAS 04232+1436 & 0.77 & -0.07 \\
IRAS 10039-3338\footnotemark[$*$] & 0.15 & 0.74 \\
IRAS 10190+1322 & 0.82 & -0.06 \\
IRAS 10565+2448 & 0.91 & 0.04 \\
IRAS 13001-2339 & 0.71 & -0.11 \\
NGC 5256 & 0.76 & -0.09 \\
Mrk 673 & 0.67 & -0.18 \\
IRAS 14348-1447\footnotemark[$*$] & 0.80 & 0.08 \\
Mrk 848 & 0.98 & -0.01 \\
NGC 6090 & 0.98 & -0.08 \\
IRAS 17132+5313 & 0.90 & -0.04 \\
IRAS 20010-2352\footnotemark[$*$] & 0.47 & 0.07 \\
II ZW 96 & 1.03 & 0.05 \\
NGC 828 & 0.83 & -0.14 \\
UGC 5101\footnotemark[$*$] & 0.19 & 0.40 \\
NGC 4194 & -0.38 & -0.28 \\
\end{longtable}

\subsection {Global star formation relation from GMA scale to high-z galaxies}

\begin{figure}
 \begin{center}
  \includegraphics[width=7.5cm]{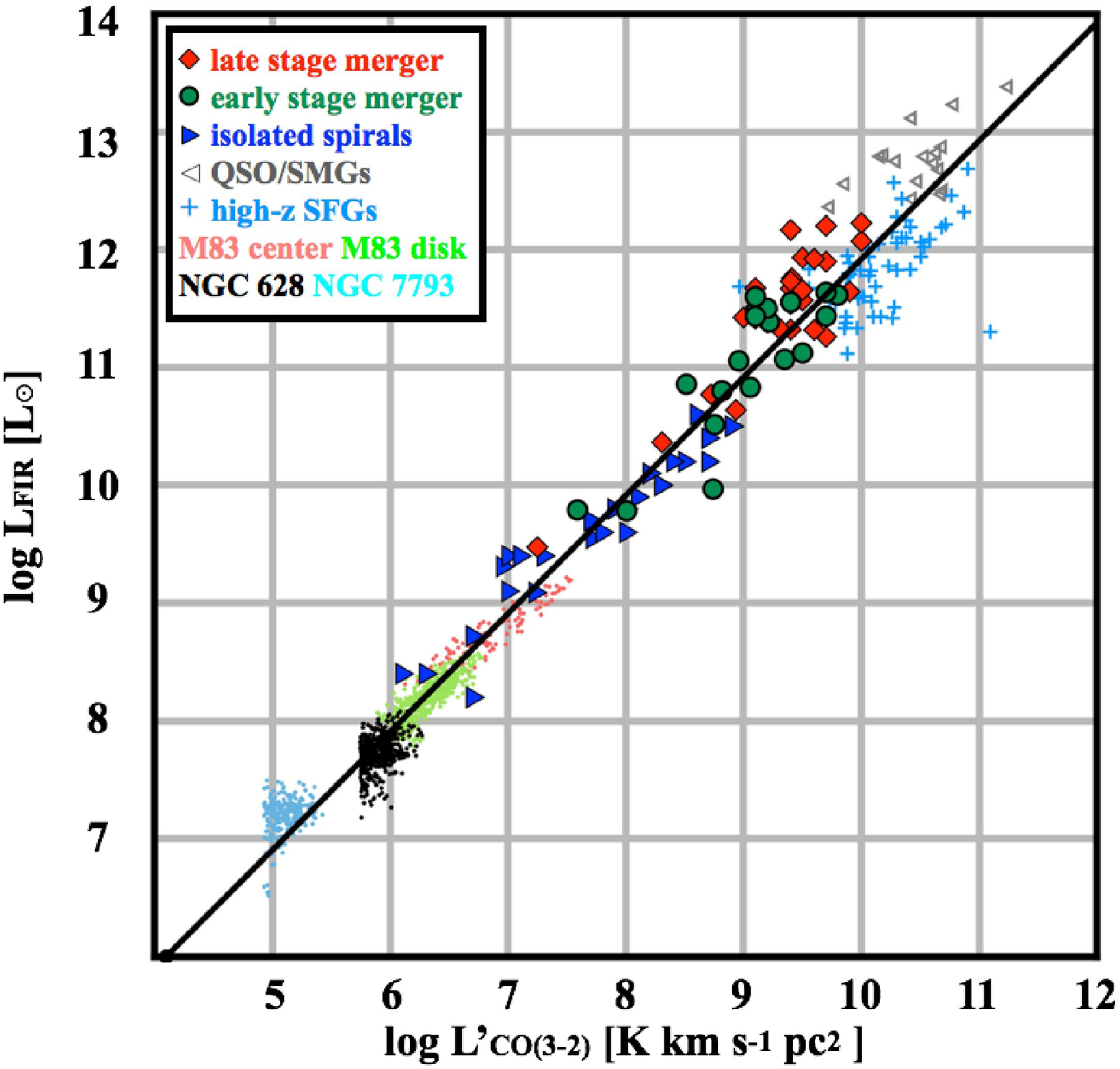} 
 \end{center}
\caption{Comparison of high-z normal star forming galaxies \citep{Tacconi+13} and SMGs/QSOs \citep{Bothwell+13,Solomon-Vanden+05} in FIR-CO plane. The GMAs in \citet{Muraoka+16} are shown as small dots. The line shows the non-weighted least-square results for all sources ($L_{\rm FIR}= 1.0 ~{\rm log}L'_{\rm CO(3-2)} + 1.0$). We note the possibility of a systematic uncertainty for deriving FIR luminosity of sample sources.
For example, we derived FIR luminosity of sample sources in \citet{Tacconi+13} from their SFR by assuming ${\rm SFR}~[\MO~{\rm yr}^{-1}]= 4.5\times 10^{44}~L_{\rm FIR}$ \citep{GFE}. In addition, the FIR luminosity of sample sources in Muraoka et al. (2016) is measured by assuming $L_{\rm IR} = 1.3L_{\rm FIR}$ \citep{Gracia+08}.}
\label{high-z}
\end{figure}

\begin{figure}
 \begin{center}
  \includegraphics[width=7.5cm]{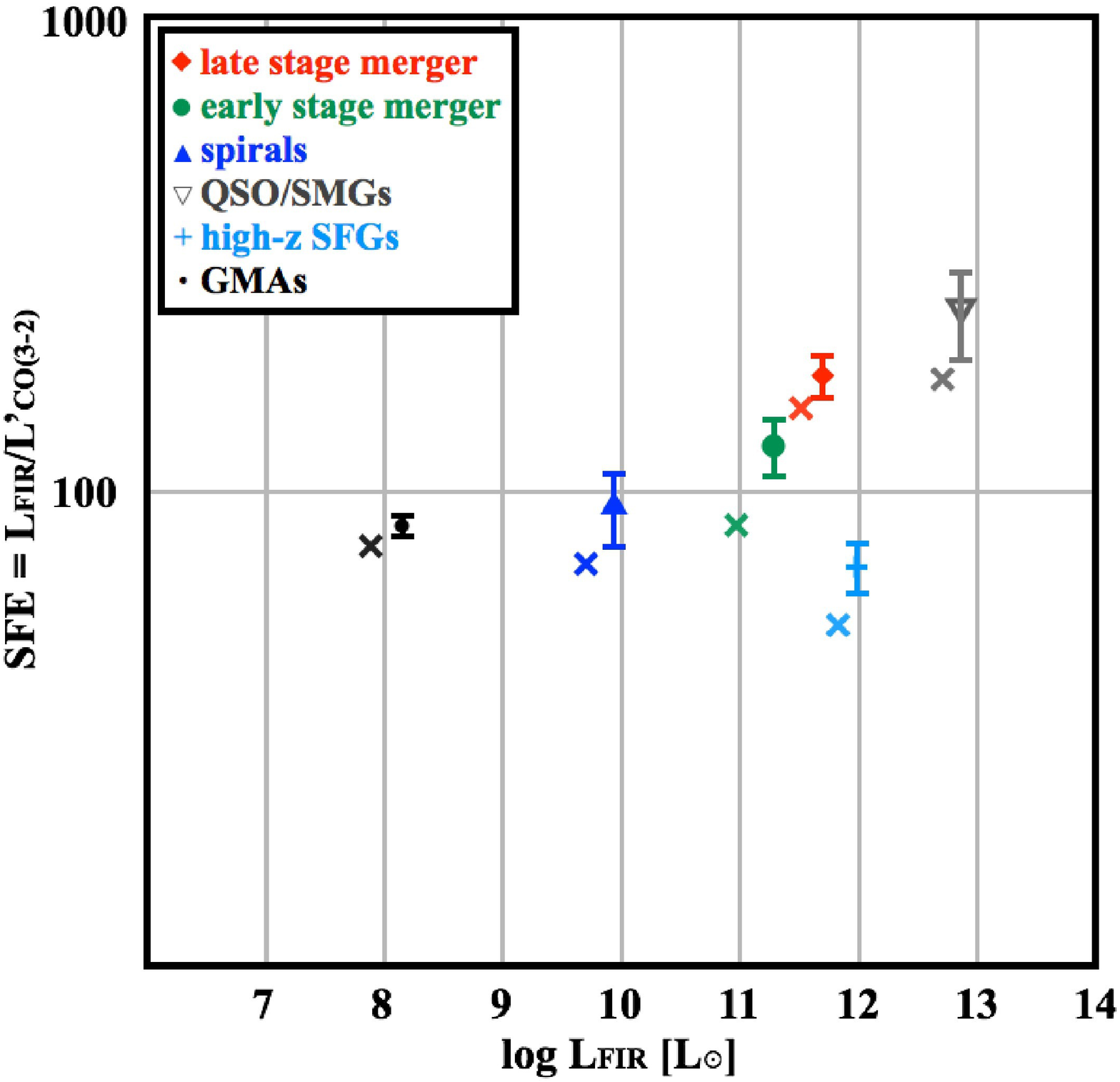} 
 \end{center}
\caption{The relation between FIR-SFE for average value (median values are plotted as crosses). The error bars are calculated as a standard error.}
\label{SFE-all}
\end{figure}

We investigate the dependence of the  $L'_{\rm CO(3-2)}$ and $L_{\rm FIR}$ relation on different star forming environments (e.g., normal star formation in isolated spirals and extreme star-burst in SMGs) using the molecular gas observations in distant galaxies \citep{Carilli+13,Casey+14}.  
We use the CO~(3--2) data from normal high-z star forming galaxies  \citep{Tacconi+13} and SMGs/QSOs \citep{Solomon-Vanden+05,Bothwell+13}.  
In addition, we use the ASTE on-the-fly CO~(3--2) images of NGC~628, NGC~7793 \citep{Muraoka+16}, and M~87 \citep{Muraoka+07} for the GMA (sub-kpc) scale star forming region.
We show the comparison with GMAs, high-z sources, and our sample sources in figure~\ref{high-z}.
We conducted statistical non-weighted least-square fitting for all sources plotted in figure~\ref{high-z}, and obtained a linear relation ($L_{\rm FIR}= 1.0 ~{\rm log}L'_{\rm CO(3-2)} + 2.0$) across six orders of magnitude in the CO~(3--2) luminosity. While the global linear relation is consistent with previous studies (e.g., \cite{Greve+14}), the relationships are different if we compare the different populations separately.
For example, high-z SMG/QSOs (gray triangles) and nearby late stage mergers (red diamonds) are systematically higher than the global fit especially seen at the galaxies with higher IR-luminosity is consistent with previous observation \citep{Gao+07} and recovered by GMCs model prediction \citep{Krumholz+07,Nara+08r}.
The high-z normal star forming galaxies \citep{Tacconi+13} are on lower side on average. 
This possibly suggests that a single global relation cannot explain the star formation from GMA scale to high-z galaxies.

In addition, we found that the average SFE of each population gradually increases from isolated galaxies, merging galaxies, and to high-z SMG/QSOs (figure \ref{SFE-all}). This may suggest that high-z SMG/QSOs experience on-going efficient star formation possibly due to galaxies involved in a major merger. In contrast, the average SFE of high-z normal star forming galaxies (light blue) are small compared to other populations, suggesting a longer lasting mode of star formation than nearby isolated spiral galaxies. Finally, we note that the detailed mechanism for triggering active star-bursts between low and high redshifts may be different, since the gas mass fraction is expected to evolve as a function of redshift \citep{Scoville+14} and global instabilities may dominate the star-burst activities in distant gas rich sources. 

\subsection {Merger induced star formation and SFE}
The increase of SFE from isolated spiral, early stage mergers to late stage mergers can be interpreted in the context of simulated gas rich merging galaxies.
These simulations predicted that, in addition to the global gas inflow to their nuclei and central star-busts which dominate the activity in the late stage mergers, disk wide star-bursts occur early in the tidal interaction due to the exacerbated fragmentation of dense gas \citep{Teyssier+10,B+11}. This is consistent with the observational results of dense gas tracers (e.g., HCN and HCO$^{+}$) toward interacting systems, which have shown that dense gas is ubiquitous across the merging interface between the two interacting galaxies (e.g., \cite{Iono+13,Saito+15}). 
A recent similar simulation \citep{Powell+13} suggested that all merging galaxies (not only star-burst systems like the Antennae galaxy but also lesser degree ones) are in between the disk sequence and the star-burst sequence.
Assuming that the formation of disk-wide dense clumps and associated star-bursts occur early in the evolution, relatively lower SFEs in the early stage than the late stage mergers suggest that such star-bursts are relatively inefficient.
This is a sensible interpretation as the gas kinematics is still dominated by disk rotation at this stage, and localized clumpy star-bursts could consist only a small fraction of the disk surface.
In the late stages of the merger when the two nuclei are about to merge, significant amount of material is expected to be deposited into the nuclei, where the surface density of gas can be much higher and star-bursts can be more efficient than the localized clumps formed early in the evolution. 
The observed higher SFE seen in the late stage mergers is consistent with this scenario.
A systematic high resolution survey of diffuse and dense gas tracers in order to resolve clumps in the disks and nuclei are the key to confirm this scenario.

Finally, we note that ideally one should use the $M_{\rm H_2}$ and SFR relation instead of the relation between $L'_{\rm CO(3-2)}$ and $L_{\rm FIR}$ used in this study. However, the CO-to-H$_2$ conversion factor is different for star-bursting galaxies and spirals (e.g., \cite{Solomon-Vanden+05,Bolatto+13}), and the exact conversion factor may vary from source to source. On the other hand, \citet{Nara+12} show that the conversion factor varies smoothly during the course of a merger evolution. In addition, the $L_{\rm FIR}$-to-SFR conversion is reasonable for star-burst galaxies but may not be applicable for galaxies with small FIR luminosity, in which case the UV luminosity may be a better indicator of the SFR. In recent study, \citet{Bournaud+15} suggest that low CO-to-H$_2$ conversion factor ($\sim2$ $\MO$ (K kms$^{-1}$ pc$^{2}$)$^{-1}$) for star-bursting mergers due to strong ISM turbulence in SB mergers. On the other hand, their model predict that higher CO-to-H$_2$ conversion factor ($\sim4$) for high-z disk galaxies than star-burst mergers.
Nevertheless, our current data is insufficient to address this issue at present and deferred to future studies that include the analysis of the CO~(1--0) data in the same sources.

\section {Summary}

We have observed CO~(3--2) emission toward 19 early and 7 late stage merging galaxies with ASTE. 
By including galaxies that are observed in previous studies, we compare the relation between the CO~(3--2) and FIR luminosity of 29 early stage merging galaxies, 31 late stage merging galaxies, and 28 isolated spiral galaxies. 
We summarize the main findings as follows;
\begin{itemize}

 \item The CO~(3--2) luminosity and FIR luminosity are correlated in the case of both spirals and mergers (figure \ref{FIR-CO}). We do not see an obvious bimodality in the $\log L'_{\rm CO(3-2)}$--$\log L_{\rm FIR}$ plane, which was previously suggested by e.g., \citet{Daddi+new,Daddi+old,Genzel+10}.
 
 \item We suggest two different slopes between spiral galaxies and merging galaxies. The slope in the $\log L'_{\rm CO(3-2)}$--$\log L_{\rm FIR}$ plane is different between the isolated spiral galaxies ($\alpha\sim 0.79$) and merging galaxies ($\alpha\sim 1.12$) (figure \ref{slope}).
 
  \item We see a modest positive correlation between FIR luminosity and SFE in merging galaxies. This correlation is not seen in previous observation of merging galaxies (e.g, \cite{Iono+09}) due to lack of the FIR faint sources.
  
  \item The average SFE gradually increases from isolated spiral, early stage mergers to late stage mergers.   
 
 \item The possible scenario to explain our results is (1) inefficient star-bursts triggered by disk-wide dense clumps occured in the early stage of interaction and (2) efficient star-bursts triggered by central concentration of gas occur in the final stage. A systematic high resolution survey of diffuse and dense gas is key to confirm this scenario.
 
\end{itemize}
  
\bigskip
\begin{ack}
We greatly appreciate the feedback offered by Dr. Desika Narayanan, who reviewed our manuscript.\\
D.I., K.K., S.K., and T.M are supported by JSPS KAKENHI Grant Number 15H02074.\\
T.S. is financially supported by a Research Fellowship from the Japan Society for the Promotion of Science for Young Scientists.\\
The ASTE telescope is operated by National Astronomical Observatory of Japan (NAOJ).\\
This research is based on observations with {\itshape AKARI}, a JAXA project with the participation of ESA.\\
This publication makes use of data products from the Wide-field Infrared Survey Explorer, which is a joint project of the University of California, Los Angeles, and the Jet Propulsion Laboratory/California Institute of Technology, funded by the National Aeronautics and Space Administration.\\
This research has made use of the NASA/IPAC Extragalactic Database (NED) which is operated by the Jet Propulsion Laboratory, California Institute of Technology, under contract with the National Aeronautics and Space Administration.
The National Radio Astronomy Observatory is a facility of the National Science Foundation operated under cooperative agreement by Associated Universities, Inc.
\end{ack}

{}

\end{document}